%% file: 2016_tcfd_KAISER.tex
\RequirePackage{fix-cm}
\documentclass[smallcondensed,final]{svjour3}     % onecolumn (ditto)
\smartqed  % flush right qed marks, e.g. at end of proof
\usepackage{graphicx}
\usepackage{amsmath,amssymb,subfigure,psfrag}
\usepackage{xcolor}

\journalname{Theor. Comput. Fluid Dyn.}

\graphicspath{{.}{Figures/}}

\usepackage[verbose,lmargin=2.75cm,rmargin=2.75cm,tmargin=3.25cm,bmargin=3.cm,marginpar=0pt,marginparsep=0pt]{geometry}
\begin{document}

\title{Cluster-based control of nonlinear dynamics%\thanks{Grants or other notes
}
% \subtitle{Do you have a subtitle?\\ If so, write it here}


\author{Eurika Kaiser         \and
        Bernd R. Noack        \and 
        Andreas Spohn         \and 
        Louis N. Cattafesta   \and 
        Marek Morzy\'nski%etc.
}


\institute{E. Kaiser \at
              University of Washington, Mechanical Engineering Department, Seattle, WA 98195, USA\\
              \email{eurika.kaiser@gmail.com}  
            \and
           E. Kaiser \and B.~R. Noack \and A. Spohn \at
              Institut PPRIME, UPR 3346 CNRS -- Universit\'e de Poitiers -- ENSMA,
	      F-86961 Futuroscope Chasseneuil, France
           \and 
           B.~R. Noack \at
              LIMSI-CNRS, UPR 3251, F-91405 Orsay cedex, France, France\\
              Technische Universit\"at Braunschweig, D-38108 Braunschweig, Germany
           \and 
           L.~N. Cattafesta \at
              Florida State University, Florida Center for Advanced Aero-Propulsion, Tallahassee, FL 32310, USA
           \and
           Marek Morzy\'nski \at
	      Poznan University of Technology, Piotrowo 3, 60-965 Poznan, Poland
}

\date{Received: date / Accepted: date}

\maketitle

\begin{abstract}
The ability to manipulate and control fluid flows 
is of great importance in many scientific and engineering applications.
% It is often the aim to change certain average properties of the flow such as the average drag or lift.
Here, a cluster-based control framework is proposed to 
determine optimal control laws with respect to a cost function 
% deduce optimal control laws from a specified objective function
for unsteady flows.
The proposed methodology 
frames high-dimensional, nonlinear dynamics into 
low-dimensional, probabilistic, linear dynamics
which considerably simplifies the optimal control problem 
while preserving nonlinear actuation mechanisms.
The data-driven approach builds upon 
a state space discretization using a clustering algorithm
which groups kinematically similar flow states into a low number of clusters.
The temporal evolution of the probability distribution on this set of clusters
is then described by a Markov model.
% The derived Markov model is here a critical enabler for designing the optimal control law 
% for which the Markov model predicts a desired invariant probability distribution.
The Markov model can be used as predictor for the ergodic probability distribution
for a particular control law.
This probability distribution approximates the long-term behavior of the original system 
on which basis the optimal control law is determined.
% and thus can be used to compute to average flow properties based on Birkoff's ergodic theorem.
The approach is applied to a separating flow dominated by the Kelvin-Helmholtz shedding.

% For ergodic systems,
% the long-term behavior of a system can be predicted by an invariant probability distribution
% associated with the Markov model.
% Thus, by designing a control law for which the Markov model predicts a desired probability distribution 
% The Markov model is able to predict the long-term behavior of the flow in terms of a probability distribution.
% building upon a statistical description of the flow
% By assuming a probabilistic point of view of the flow,
% the 
% In particular, 
% a state space and time discretization of the classical Liouville equation is employed
% which simplifies the 
% statistical description of the flow 

\end{abstract}

\section{Introduction}
\label{Sec:Intro}
\input{S1}

\section{Cluster-based control methodology}
\label{Sec:Methodology}
\input{S2}

% \section{Control of the generalized mean-field model}
% \label{Sec:gMFM}
%\input{S3}

\section{Control of a separating flow over a smooth ramp}
\label{Sec:Ramp2u}
\input{S4}

\section{Conclusions}
\label{Sec:Discussion}
\input{S5}

\begin{acknowledgements}
The authors acknowledge the funding and excellent working conditions 
of the project ’Separation Control - From passive to closed-loop design’ (SepaCoDe, ANR-11-BS09-018), 
the Chair of Excellence ’Closed-loop control of turbulent shear flows using reduced-order models’ (TUCOROM, ANR-10-CHEX-0015), 
both supported by the French Agence Nationale de la Recherche (ANR) and hosted by Institute PPRIME,  
the Collaborative Research Center (CRC 880) 'Fundamentals of High Lift for Future Civil Aircraft' 
funded by the German Research Foundation (DFG)
and hosted at the Technical University of Braunschweig, Germany, and the 
project ``Novel Method of Physical Modal Basis Generation for Reduced Order Flow Models''
funded by the Polish National Centre of Science under research grant no. 2011/01/B/ST8/07264.
EK also thanks for the great support through the region Poitou-Charentes, the NSF PIRE Grant OISE-0968313, 
and the Air Force research lab under grant AFRL FA8651-16-1-0003.
%MS would like to acknowledge the support of the LINC
%project (no. 289447) funded by EC's Marie-Curie ITN program
%(FP7-PEOPLE-2011-ITN). 
%We thank the Ambrosys GmbH (Society for Complex Systems Management)
%and the Bernd Noack Cybernetics Foundation for additional support.
We appreciate valuable stimulating discussions
with:
Bing Brunton, 
Steven Brunton,
Eric Deem,
Nicolai Kamenzky, 
Nathan Kutz, and 
Robert Niven.
%\cek{Last but not least, we thank the referees
%for many important suggestions.}

%Special thanks are due to Nadia Maamar 
%for a wonderful job in hosting the TUCOROM visitors.
\end{acknowledgements}

\appendix
\input{appendix}

\bibliographystyle{spmpsci}      % mathematics and physical sciences
\bibliography{2016_tcfd_KAISER}   % name your BibTeX data base

% \begin{thebibliography}{}
% %
% % and use \bibitem to create references. Consult the Instructions
% % for authors for reference list style.
% % Format for Journal Reference
% Author, Article title, Journal, Volume, page numbers (year)
% Author, Book title, page numbers. Publisher, place (year)
% \end{thebibliography}

\end{document}

%% file: S1.tex
% When controling fluid flows, a deterministic viewpoint is often assumed.
% The evolution of a single trajectory is tracked and, e.g., its derivation to a reference state or trajectory shall be minimzed.
% 
% The ability to control and manipulate --> technological benefits --> many applications
Controlling complex dynamical systems such as fluid flows 
is of great importance in science and engineering.
Examples include drag reduction for greener transport systems, 
lift increase on airfoils,
stabilization of combustion processes,
reduction of pollutants from chemical processes,
and efficiency increase of energy harvesting systems like wind turbines, to name a few.
% The understanding and exploitation of nonlinear actuation mechanisms like frequency cross-talk 
% play a crucial role in developing effective control strategies. \textcolor{red}{TODO: add examples.}
Closed-loop control which translates the continuously monitored system state into control actions
is a particularly promising direction. 
We refer to \cite{Brunton2015amr} for a recent review on closed-loop control. 

Of particular interest in control applications are certain statistical flow properties 
like the average drag or lift which shall be mitigated or increased, respectively. %\textcolor{red}{TODO: add why.}
However, their computation from trajectories may be misleading for several reasons.
The time average is generally computed over a limited time span
which makes it sensitive to transient behavior and 
biased as the trajectory may reside only in a confined state space region.
Thus, very long integration times are required to ensure that the time average is good.
% Average properties computed along a trajectory 
% may not be exact as these are often based on averaging over a certain time span
% in which the trajectory may reside only in a certain state space region.
% From a practical point of view, 
% very long sampling horizons may be required which is not always realizable.
% On the other hand, 
% the long-term behavior of a time series can be unpredictable over long integration times. 
However, even without noise and external disturbances, 
small uncertainties in initial or boundary conditions may doom a deterministic system unpredictable.
A well-studied example is the chaotic Lorenz system introduced by E.~N. Lorenz \cite{Lorenz1963jas},
a simplified model for atmospheric convection with known sensitivity to initial conditions.

Average properties over long time spans %are of particular interest %generally considered 
% naturally leading 
lead naturally to invariant probability measures on the attractor, i.e.\
these measures stay the same after transformation of the attractor. 
Ergodic measures, a sub-class of invariant measures, are of particular interest 
as for those time averages are equal to space averages according to Birkhoff's ergodic theorem \cite{Lasota1994book}.
This assumption %is a strong assumption which 
is often assumed when analyzing fluid flows: 
% For the analysis of many flows, 
% these flows are assumed to be ergodic, 
These flows are assumed to be ergodic, 
i.e.\ in the sense that they are statistically reproducible,
allowing to compute the statistical properties from ensemble averages.
In this study, the system's dynamics are modelled in terms of a Markov model, 
particularly a cluster-based reduced-order model (CROM) \cite{Kaiser2014jfm}.
This simplification allows to compute many (statistical) properties exactly
which are often good estimators for the analogous properties of the original system \cite{Froyland2001}.
As a consequence of Birkhoff's ergodic theorem,
controlling such statistical properties is strongly related to the control of the ergodic measure on the attractor.

% The aim of this study is to derive optimal control laws producing desired ergodic probability distributions as predicted by CROM.
The Markov model is a linear evolution equation for a probability distribution in the state space.
Such evolution equations can be derived from the Navier-Stokes equation, 
starting with the linear Liouville equation of a suitable probability space. 
The Hopf \cite{Hopf1952jrma} formalism for the Navier-Stokes equation is a prominent example. 
A simpler version constitutes the Liouville equation for a Galerkin system. 
The reader is referred to \cite{Noack2012jfm} for a detailed discussion. 
CROM is closely aligned with closure schemes, in which a stable fixed point represents the ergodic measure 
for the unsteady attractor in velocity space.
% Control of the Lioville equations is a vividly studied field \cite{???}.
% As elaborated in \cite{???}, 
% many realizations corresponding to different initial conditions.
While the control of a Liouville equation has not found much attention in fluid dynamics yet, 
it is studied widely in other fields such as atomic physics \cite{Munowitz1987jcp}, 
biology \cite{Brockett2010proc}, and robotics \cite{Brockett2003ieee,Majumdar2014ijrr}.
An extensive study on the optimal control of the Liouville equation is provided in \cite{Brockett2012}.
% Generally, two interpretations are pursued:
% On the one hand, the control of many copies of an identical system is considered 
% such as the scattering behavior of physical particles.
% On the other hand, 
% a single controller manipulates a particular system over repeated realizations corresponding to different initial conditions.
% Thus, the control of the Liouville equation is a promising direction for systems 
% that exhibit uncertainties in initial conditions and system parameters, e.g. due to disturbances.
The control of the Liouville equation can be interpreted as 
the manipulation of a particular system using a single controller 
over repeated realizations which correspond to different initial conditions.
Thus, the control of the Liouville equation is a promising direction for systems 
that exhibit uncertainties in initial conditions and system parameters, e.g. due to disturbances.

The present work is outlined as follows:
In Sec.~\ref{Sec:Methodology}, the cluster-based control methodology is described.
The approach is applied to the benchmark problem 
of a separating flow over a backward-facing, smoothly contoured ramp
which results are presented in Sec.~\ref{Sec:Ramp2u}.
The main results are summarized and discussed in Sec.~\ref{Sec:Discussion}.
% Practical implementation details and additional theoretical concepts are provided in the appendix
% to keep the principal presentation clear and concise and 
% to provided additional information of practical importance.
Details on the empirical estimation of the Markov model and its properties
are given in Appendix~\ref{App:Sec:ClusterAnalysis}.
% Properties of the model which are crucial for the control design are provided in Appendix~\ref{App:Sec:PropertiesOfCTM}.
In Sec.~\ref{App:Sec:CLviz}, a technique for visualizing the similarity of control laws is briefly explained.

%% file: S2.tex
\subsection{Problem formulation}
In this work, we are concerned with identifying a probabilistic low-order representation of the deterministic, fully nonlinear dynamics
and deriving optimal control laws with respect to an objective function.
Generally, a dynamical system is represented as
% ------------ Equation -------------- %
\begin{eqnarray}
 \frac{\mathrm d}{\mathrm dt}\vec{a}(t) &=& \vec{F}(\vec{a}(t), \vec{b}(t))%\\
%  \vec{s}(t) = \vec{C}\vec{a} + \vec{D}\vec{b}
 \label{Eqn:NonlinearSystem}
\end{eqnarray}
% ------------------------------------ %
where the vector $\vec{a}$ %\in\mathcal{A}\subset \mathbb{R}^n$ 
denotes the system state and vector $\vec{b}$ %\in\mathcal{B}\subset \mathbb{R}^m$ 
is the control input at time $t$, and $\vec{F}$ is the nonlinear propagator for the system state $\vec{a}$.
We assume a full-state feedback ansatz for the control in the form
% ------------ Equation -------------- %
\begin{equation}
 \vec{b} = K(\vec{a})
\end{equation}
% ------------------------------------ %
where $K$ represents the control law that maps states $\vec{a}$ into control actions $\vec{b}$.
In optimal control, one seeks to determine an optimal control law $K^{opt}$ 
which minimizes a cost function.
% that is generally a function of the system state $\vec{a}$ and the control $\vec{b}$.
The cost function, 
generally a function of the system state $\vec{a}$ and the control $\vec{b}$, 
defines the control objective through a performance measure and penalty function 
evaluating the cost of the applied control,
and incorporates additional constraints.
% The unsteady behavior and inherent nonlinearities of fluid flows often pose severe challenges 
% in deriving effective feedback control strategies.
% Due to the unsteady and inherently nonlinear beahvior of many fluid flows and disturbances,
% one is generally interested in changing mean properties of the flow, e.g. the average drag or lift force.
In flow control, the average drag or lift are often of interest.
Let $h(\vec{a})$ be a function measuring a quantity of interest,
e.g., the drag, along the trajectory trajectory $\vec{a}(t)$.
For ergodic behavior, the temporal average 
$\langle h(\vec{a}) \rangle_T := \lim_{T\rightarrow\infty}\int_{0}^T \, h(\vec{a}(t))\,\mathrm dt$ 
% of a function $h(\vec{a})$, e.g., a measurement of the drag, along a particular trajectory $\vec{a}(t)$
can be represented in terms of the spatial average 
$\langle h(\vec{a}) \rangle_{\Omega} := \int_{\Omega} \, h(\vec{a})\,p(\vec{a})\,\mathrm d\vec{a}$
which is naturally defined by the probability density function (p.d.f.) $p(\vec{a})$.
The average cost function can then be formulated as
% ------------ Equation -------------- %
\begin{equation}
 J_{K} 
%  = \lim\limits_{T\rightarrow\infty}\,\mathbb{E}\left[ \sum\limits_{t=0}^T h_t(\vec{a}_t, \vec{b}_t)\right]
%  = \lim\limits_{T\rightarrow\infty}(\vec{p}^{t})^T \hat{\vec{h}}. 
   = \mathbb{E}^{\infty} \left[ j(\vec{a},\vec{b})\vert_{\vec{b} = K(\vec{a})} \right]
   = \int\limits_{\Omega}\,j(\vec{a},\vec{b})\vert_{\vec{b} = K(\vec{a})}\,p^{\infty}(\vec{a})
   \,\mathrm d\vec{a}
\end{equation}
% ------------------------------------ %
where $j(\vec{a},\vec{b})$ %\in\mathcal{J}\subset \mathbb{R}$ 
is the local cost function, 
$\mathbb{E}^{\infty}$ is the expectation operator assuming transients have decayed,
%% , statistically stationary case, no transients
% evaluating the controlled system state $\vec{a}$ with respect to a control objective
% and penalizing the control input $\vec{b}$ 
and $p^{\infty}(\vec{a})$ is the asymptotic, i.e.\ long-run, p.d.f.
The control design task is to determine $K^{opt}$ such that the p.d.f.  
$p^{\infty}(\vec{a})$ is as close as possible to a desired density
for which the average cost $J_{K}$ is minimized.
%, and thus minimizing the average cost.
% In ergodic systems, 
% the temporal mean of a property can be expressed 
% Moreover, it is often only possible to predict the effect of an control for a short time 
% and the long-term behavior 
The evolution of the p.d.f. is prescribed by a Liouville equation
associated with the dynamical system \eqref{Eqn:NonlinearSystem},
% ------------ Equation -------------- %
\begin{equation}
 \frac{\partial}{\partial t}p(\vec{a},t) + \nabla_{\vec{a}}\cdot\left[p(\vec{a},t)\,
 %\left( 
 \vec{F}(\vec{a}, \vec{b})
 %\right)
 \right] = 0.
 \label{Eqn:LiouvilleEquation}
\end{equation}
% ------------------------------------ %
While the dynamical system \eqref{Eqn:NonlinearSystem} prescribes the evolution of 
a single trajectory in the state space, the Liouville equation \eqref{Eqn:LiouvilleEquation} 
is a linear equation for the p.d.f. describing the distribution of a swarm of trajectories in the state space.
Linked to the Liouville equation \eqref{Eqn:LiouvilleEquation} is the Perron-Frobenius operator $P_t$ \cite{Lasota1994book},
a linear evolution operator, that maps the p.d.f. forward in time,
% ------------ Equation -------------- %
\begin{equation}
 p(\vec{a},t) = P_t\,p(\vec{a}(0))   
 \label{Eqn:PerronFrobenius}
\end{equation}
% ------------------------------------ %
with $P_t := \exp(t\,L)$ where $L:=-\nabla_{\vec{a}}\cdot(p(\vec{a},t)\,\vec{F}(\vec{a}, \vec{b}))$ is the Liouville operator.
An invariant (or long-term) p.d.f. $p^{\infty}(\vec{a})$ constitutes a solution to the fixed-point equation $p(\vec{a}) = P_t\,p(\vec{a})   \text{ for all } t\geq 0$.
% ------------ Equation -------------- %
% \begin{equation}
%  p(\vec{a}) = P_t\,p(\vec{a})   \text{ for all } t\geq 0
%  \label{Eqn:PerronFrobeniusInvariance}
% \end{equation}
% ------------------------------------ %
% Note that this is \textit{a} particular invariant p.d.f.
Note that a unique solution is not expected and 
there can be many or even infinitely many invariant p.d.f.s.
For instance, 
if the dynamical system \eqref{Eqn:NonlinearSystem} possesses a fixed point $\vec{a}^{\star}$, 
the invariant density will be a peak supported over the fixed point,
i.e.\ $p^{\infty}(\vec{a}) = \delta(\vec{a}^{\star})$ where $\delta$ is the Dirac delta function.
If  \eqref{Eqn:NonlinearSystem} exhibits a periodic limit cycle, 
the invariant density is the sum of delta functions supported over the points $\vec{a}^{\star}_i$, $i=1,\ldots,N_{lc}$,
constituting the limit cycle,
i.e.\ $p^{\infty}(\vec{a}) = \sum_{i=1}^{N_{lc}}\delta(\vec{a}^{\star}_i)$.
If \eqref{Eqn:NonlinearSystem} is a chaotic dynamical system, 
it consists of infinitely many limit cycles and therefore of infinitely many invariant densities.
the reader is referred to \cite{Bollt2013book} for more details on this topic.

% TODO
% \begin{itemize}
%  \item transients
%  \item globally stable? will all initial distributions settle to this one
%  \item interested in the dominant one --< dominant behavior
% \end{itemize}

% The time average is generally computed over a certain limited time span
% which makes it sensitive to transient behavior and 
% biased as the trajectory may reside only in a confined state space region.
% Hence, very long integration times are required to ensure that the time average is good.
% % Average properties computed along a trajectory 
% % may not be exact as these are often based on averaging over a certain time span
% % in which the trajectory may reside only in a certain state space region.
% In this study, we aim to model the dynamics in terms of a discrete-state, discrete-time Markov model.
% This is based on a finite-rank approximation of the Perron-Frobenius operator \eqref{Eqn:PerronFrobenius}.
% This simplification allows to compute many properties exactly
% which are often good estimators for the properties of the original system 
% \textcolor{red}{(Froyland, Extracting dynamical behaviour via Markov models)}.
% sampling

\subsection{Discrete coarse-graining of state space}
Let $\mathcal{A}_i$, $i=1,\ldots,N_a$, be a discretization of the state space
such that $\mathcal{A} = \cup_{i=1}^{N_a} \mathcal{A}_i$ with  $\mathcal{A}_i \cap \mathcal{A}_j = \emptyset$ for $i\neq j$.
% A common discretization involves a partition of the state space into evenly-spaced boxes. 
% While our approach is not restricted to this choice, 
Here, a data-driven partitioning method is pursued as outlined in App.~\ref{App:Sec:ClusterAnalysis}.
% This clustering algorithm
which yields a discrete number of clusters $\mathcal{A}_i$ with centroids $\vec{A}_i$ 
which are the 
as representative states of each cluster. 
% finite-state machine/deterministic finite automaton (also time-discrete)? , 
% now 
Each state $\vec{a}(t)$ is connected to a symbol $\alpha$ representing the cluster $\mathcal{A}_{\alpha}$ to which $\vec{a}(t)$ belongs.
Let the measurable equation, which maps the continuous state $\vec{a}$ to a discrete symbol $\alpha$, be defined by 
% ------------ Equation -------------- %
\begin{eqnarray}
 \alpha (t) = \chi(\vec{a}(t)) \in \{1,2,\ldots,N_a \}.
\label{Eqn:DiscreteStateMeasurementEqn}
\end{eqnarray}
% ------------------------------------ %
% where the symbol $\alpha$ represents the clusters $\mathcal{A}_i$ and their respective centroids $\vec{A}_i$, $i=1,\ldots,N_a$.
The coarse-grained inverse mapping is defined by
% ------------ Equation -------------- %
\begin{eqnarray}
 \vec{a}^{\circ}(t) = \vec{A}_{\chi(\vec{a}(t))}
\label{Eqn:DiscreteStateMeasurementEqnInversion}
\end{eqnarray}
% ------------------------------------ %
% Furthermore, each state $\vec{a}$ is approximated by the respective cluster centroid 
approximating the continuous state $\vec{a}$ by its closest cluster centroids, 
e.g., $\alpha=1$ and $\vec{a}^{\circ} =  \vec{A}_1$ if $\vec{a}\in\mathcal{A}_1$. 
The superscript $^{\circ}$ refers to the discrete-state approximation. 
The inverse operation is associated with a loss of information due to the coarse-graining process.
Let the characteristic function be defined by 
% ------------ Equation -------------- %
\begin{equation}
 \chi_i(\vec{a}) = \begin{cases}
                           1 & \text{if } \vec{a}\in\mathcal{A}_i,\\
                           0 & \text{if } \vec{a}\notin\mathcal{A}_i
                          \end{cases}    
                          \quad\text{or}\quad
 \chi_i(\vec{a}) =  \delta(\chi(\vec{a})-i)                        
\label{Eqn:IndicatorFunction}
\end{equation}
% ------------------------------------ %
where $\delta$ is the Kronecker delta.
% Note that in contrast to \eqref{Eqn:IndicatorFunction}, 
% $\chi(\vec{a}(t))$ in \eqref{Eqn:DiscreteStateMeasurementEqn} 
% Note that in contrast to $\chi_i(\vec{a})$,
% $\chi(\vec{a}(t))$
% assumes the respective symbol of the prevailing cluster as value.
The full-state feedback ansatz for the control is then
% ------------ Equation -------------- %
\begin{equation}
% \begin{subequations}
%   \begin{align}
 \vec{b}^{\circ} = K(\vec{a}^{\circ}) = K\left(\vec{A}_{\chi\left(\vec{a}\right)}\right)
%        b &=& \kappa(\alpha) = \kappa(\chi(\vec{a})). 
           = \kappa(\alpha) = \sum\limits_{i=1}^{N_a}\, \vec{B}_i\, \chi_{i}(\vec{a}). 
%   \end{align}
% \end{subequations}
\end{equation}
% ------------------------------------ %
As a result of the discretization, the control $\vec{b}$ is piecewise constant
where $\vec{B}_i$ are vectors of real numbers and denote the control applied in cluster $\mathcal{A}_i$.
% The control input may also be a scalar $B_i$ in the case of single-input control.
% For instance, $\vec{b}^{\circ} = \vec{B}_1$ for $\vec{a}\in \mathcal{A}_1$ or $\alpha=1$.
The control law $\kappa$, 
that maps discrete states $\vec{A}_i$ with symbols $\alpha$ into control actions $\vec{b}^{\circ}$,
is considered stationary here, 
i.e.\ %a particular value 
$\vec{B}_{\alpha}$ for a cluster $\alpha$ remains constant for all times.
The optimal control law $\kappa^{opt}$ minimizes the average cost function
% ------------ Equation -------------- %
\begin{equation}
 J_{\kappa} 
%  = \lim\limits_{T\rightarrow\infty}\,\mathbb{E}\left[ \sum\limits_{t=0}^T h_t(\vec{a}_t, \vec{b}_t)\right]
%  = \lim\limits_{T\rightarrow\infty}(\vec{p}^{t})^T \hat{\vec{j}}. 
   = \mathbb{E}^{\infty} \left[ j^{\circ}(\alpha,\vec{b})\vert_{\vec{b}=\kappa(\alpha)} \right]
   = \sum\limits_{i=1}^{N_a}\,
   j^{\circ}(i,\kappa(i)) \;p^{\circ,\infty}_i
\end{equation}
% ------------------------------------ %
with the local cost function $j^{\circ}(\alpha,\vec{b})\vert_{\vec{b}=\kappa(\alpha)}$ 
evaluating the cost for the currently prevailing cluster $\alpha$ and the control input $B_{\alpha}$ applied in this cluster.
The vector $\vec{p}^{\circ,\infty}$ is the discrete asymptotic probability distribution for $t\rightarrow\infty$.
This constitutes the solution to the fixed-point equation 
%$\vec{p}^{\infty} = \vec{P}_{\kappa}\,\vec{p}^{\infty}$
associated with 
the discrete-state Markov model
% ------------ Equation -------------- %
\begin{equation}
 \frac{\mathrm d}{\mathrm dt}\vec{p}^{\circ}(t) = \tens{P}_{\!\kappa}^{\circ}\,\vec{p}^{\circ}(t).
 \label{Eqn:DiscreteStateMarkovModel}
\end{equation}
% ------------------------------------ %
This equation describes the temporal evolution of the probability vector 
$\vec{p}^{\circ} = [p_1^{\circ},\ldots,p_{N_a}^{\circ}]^T$ where $p_i$ is the probability that the trajectory $\vec{a}(t)$
resides in cluster $\mathcal{A}_i$.
The matrix $\tens{P}_{\!\kappa}^{\circ}$ prescribes the dynamics on the coarse-grained state space 
following a particular control law $\kappa$.

% It can be shown \cite{Li1976jat}
The Markov model \eqref{Eqn:DiscreteStateMarkovModel} 
can be derived from \eqref{Eqn:LiouvilleEquation} using Ulam's method \cite{Ulam1964book, Li1976jat} 
which is classically used in dynamical systems to determine a
finite-rank approximation of the Perron-Frobenius operator \eqref{Eqn:PerronFrobenius}.
Ulam's method involves a Galerkin projection of the Liouville equation \eqref{Eqn:LiouvilleEquation} onto a particular set of basis functions.
Recently, 
\cite{Kaiser2014jfm} showed that the cluster-based reduced-order modeling approach can be interpreted as 
a generalization of Ulam's method.

% Equation constitutes the discrete-state formulation of the Liouville equation.

\subsection{Discrete-time, discrete-state formulation}
\label{Sec:DSDC-Formulation}
A further discretization level based on the time is considered.
Let the 
floor function be defined as 
$\lfloor t \rfloor := \max \{l\in\mathbb{Z}\;\vert\; l\leq t \}$ \cite{Iverson1962book}. 
%(before Gaussian brackets).
A polymorphism for continuous-time and discrete-time variables is pursued for the purpose of a better readability.
The measurement equation in analogy to \eqref{Eqn:DiscreteStateMeasurementEqn} is given by
% ------------ Equation -------------- %
\begin{eqnarray}
\alpha^t = \chi(\vec{a}(\lfloor t \rfloor)) \in \{1,2,\ldots,N_a \}
%  \label{Eqn:NonlinearSystem}
\end{eqnarray}
% ------------------------------------ %
for discrete times $t$, the superscript $^t$ is an index corresponding to multiple of $\Delta t$,
and where $\chi$ is the characteristic function as introduced above.
The inverse discrete-time, coarse-grained mapping is given by
% ------------ Equation -------------- %
\begin{eqnarray}
 \vec{a}^{\bullet,t} = \vec{A}_{\chi(\vec{a}(\lfloor t \rfloor))}.
% \label{Eqn:NonlinearSystem}
\end{eqnarray}
% ------------------------------------ %
The superscript $^{\bullet}$ refers here to the discrete-state, discrete-time representation of quantities.
The full-state feedback ansatz for the control becomes
% ------------ Equation -------------- %
\begin{eqnarray}
 \vec{b}^{\bullet} &=& K(\vec{a}^{\bullet}) = K\left(\vec{A}_{\chi(\vec{a}(\lfloor t \rfloor))}\right) = \kappa(\alpha)
\end{eqnarray}
% ------------------------------------ %
realizing the time-delay.
% and where $\kappa$ represents the control law that maps discrete states $\alpha$ into control actions $\vec{b}$.
The optimal control law $\kappa^{opt}$ shall minimize the average cost function
% ------------ Equation -------------- %
\begin{equation}
 J_{\kappa} 
%  = \lim\limits_{T\rightarrow\infty}\,\mathbb{E}\left[ \sum\limits_{t=0}^T h_t(\vec{a}_t, \vec{b}_t)\right]
%  = \lim\limits_{T\rightarrow\infty}(\vec{p}^{t})^T \hat{\vec{h}}. 
   = \mathbb{E}^{\infty} \left[ j^{\bullet}(\alpha,\vec{b})\vert_{\vec{b}=\kappa(\alpha)} \right]
   = \sum\limits_{i=1}^{N_a}\,
   j^{\bullet}(i,\kappa(i)) \,p_i^{\bullet,\infty}
   \label{Eqn:DiscreteStateDiscreteTimeAveragedCost}
\end{equation}
% ------------------------------------ %
with local cost function $j^{\bullet}(\alpha,\vec{b})\vert_{\vec{b}=\kappa(\alpha)}$. %\in\mathcal{J}^{\bullet}\subset \mathbb{R}^{N_c}$.
The asymptotic probability vector $\vec{p}^{\bullet,\infty}$ 
is a solution to the fixed-point equation of the discrete-state, discrete-time Markov model
which describes consecutive distributions by the iteration formula
% ------------ Equation -------------- %
\begin{equation}
 \vec{p}^{\bullet,t+1} = \tens{P}_{\kappa}^{\bullet}\,\vec{p}^{\bullet,t},\quad t=0,1,2,\ldots
  \label{Eqn:DiscreteStateDiscreteTimeMarkovModel}
\end{equation}
% ------------------------------------ %
with the cluster transition probability matrix 
$\tens{P}_{\kappa}^{\bullet}$ prescribing the dynamics following a particular control law $\kappa$.

% The discrete-time Markov model \eqref{Eqn:DiscreteStateDiscreteTimeMarkovModel}
% can be derived from \eqref{Eqn:DiscreteStateMarkovModel} 
% by an analytical integration over the time $\Delta t$.
% The resulting discrete-time transition matrix reads $\tens{P}^{\bullet} = \exp(\tens{P}^{\circ}\,\Delta t)$.
% Then, $\tens{P}^{\circ} = \frac{1}{\Delta t}\,\log(\tens{P}^{\bullet}) \approx \frac{\tens{P}^{\bullet}-\tens{I}}{\Delta t} + \mathcal{O}(\Delta t)$
% with identity matrix $\tens{I}$
% which coincides with the Euler integration scheme for \eqref{Eqn:DiscreteStateMarkovModel}. 
% Matrix $\vec{P}_{\kappa}$ can be empirically determined as outlined in App.~\ref{App:Sec:coCROM}.

% Its spectral property are important for the control design and 

% Is only data available 
% It is commonly accepted in the Ulam–Galerkin framework that P jk cont
% can be approximated by
% can be derived from \eqref{Eqn:LiouvilleEquation} using Ulam's method \cite{Li1976}
% which is classically used in dynamical systems to determine a
% finite-rank approximation of the Perron-Frobenius operator.
% Recently, 
% \cite{Kaiser2014jfm} showed that the cluster-based reduced-order modeling approach can be interpreted as 
% a generalization of Ulam's method.

\subsection{Control design using CROM}
% \subsubsection{Infinite-horizon control using spectral analysis}
In the following, the discrete-state, discrete-time formulation of Sec.~\ref{Sec:DSDC-Formulation} is considered 
and the superscript $^\bullet$ is dropped.
We consider single-output control laws of the form
% ------------ Equation -------------- %
\begin{equation}
 \kappa(\alpha) = \sum\limits_{i=1}^{N_a}\, B_i\, \chi_{i}(\vec{a}) 
 = \sum\limits_{i=1}^{N_a}\, \tilde{\kappa}\,\sin(\omega_p\,t)\,\chi_{i}(\vec{a})
 \label{Eqn:ControlLaw}
\end{equation}
% ------------------------------------ %
where $\tilde{\kappa}$ denotes a fixed amplitude for the actuation, 
the cluster-dependent control values become scalars $B_1 = B_2 = \ldots = B_{N_a} = \tilde{\kappa}\,\sin(\omega_p\,t)$,
and $\chi_{i}(\vec{a})$ is the characteristic function defined in \eqref{Eqn:IndicatorFunction}.
As $\chi_{i}(\vec{a})$ assumes values $0$ or $1$,
the periodic actuation is turned off or on, respectively, depending on the prevailing cluster $\alpha=i$.
Thus, the control law $\kappa(\alpha)$ is uniquely determined by the characteristic function $\chi_{i}(\vec{a})$.

Let be assumed that the state space is discretized into $N_a=10$ clusters.
Since $\chi_{i}(\vec{a})$ can only assume two possible values
and the number of clusters is fixed, 
there exists a fixed number of possible control laws defined by 
all possible combinations of '0's and '1's. 
Let the control law be represented by a string of '0's and '1's of length $N_a$.
The number of '1's in this string shall be denoted as $N_v$.
% The number of ways how $N_v$ '1's can be arranged in this string, i.e.\ on $N_a$ clusters, is given by
% %combinations of $N_v$ '1's among $N_a$ clusters is given by
% %n things taken k at a time
% % ------------ Equation -------------- %
% \begin{equation}
%  C(N_a,N_v) = \binom{N_a}{N_v} = \frac{N_a!}{N_v!(N_a-N_v)!}.
% \end{equation}
% % ------------------------------------ %
% Then, the number of $N_v$-combinations for all $N_v$ is given by
The total number of combinations of how $N_v$ '1's can be arranged in this string, i.e.\ on $N_a$ clusters, is given by 
$\sum_{0\leq N_v\leq N_a}\, C(N_a,N_v) = \sum_{0\leq N_v\leq N_a}\, \binom{N_a}{N_v} = 2^{N_a}$.
% ------------ Equation -------------- %
% \begin{equation}
%  \sum\limits_{0\leq N_v\leq N_a}\, C(N_a,N_v) = \sum\limits_{0\leq N_v\leq N_a}\, \binom{N_a}{N_v} = 2^{N_a}.
% \end{equation}
% ------------------------------------ %
For the given example of $N_a=10$ clusters, 
the total number of control laws is thus $\sum_{0\leq N_v\leq 10}\, \binom{10}{N_v} = 2^{10} = 1024$:
% we have a list of $2^{10}=1024$ control laws:
% ------------ Equation -------------- %
\begin{subequations}
 \begin{align}
 \kappa^{0}(\alpha) &= B_{0000000000} = 0,\\
 \kappa^{1}(\alpha) &= B_{0000000001} = B_1\,\delta(\alpha-1),\\
%  \kappa^{2}(\alpha) &= B_{0000000010} = B_2\,\delta(\alpha-2),\\
 &\vdots\notag\\
 \kappa^{386}(\alpha) &= B_{0110000010} = B_2\,\delta(\alpha-2) + B_8\,\delta(\alpha-8) + B_9\,\delta(\alpha-9)\label{Eqn:ControlLaw386},\\
 &\vdots\notag\\
 \kappa^{1023}(\alpha) &= B_{1111111111} = \sum_{i=1}^{N_a}\,B_{i}\,\delta(\alpha-i) .
 \end{align}
\end{subequations}
% ------------------------------------ %
where $B_{{\sf xxxxxxxxxx}}$ with ${\sf x}\in\{0,1\}$ refers to the string representing the control law.
% where $B_1 = B_2 = \ldots = B_{N_a} = \tilde{\kappa}\,\sin(\omega_p\,t)$ and $\delta(\alpha-i)$ the Kronecker delta. 
% Note that counting starts at zero. 
The control design task consists of determining the control law that 
minimizes the average cost function \eqref{Eqn:DiscreteStateDiscreteTimeAveragedCost}.
For any control law $\kappa(\alpha)$ (or $\kappa^l(\alpha)$, respectively) such a cost $J_{\kappa}$ can be simply evaluated,  
% ------------ Equation -------------- %
\begin{equation}
 J_{\kappa} = \sum\limits_{i=1}^{N_a}\, j(i,\kappa(i))\,p_{\kappa,i}^{\infty} 
            = \sum\limits_{i=1}^{N_a}\, j(i,\kappa(i))\, p_{\kappa,i}^{*1},
 \label{Eqn:AverageCostEigenvector}
\end{equation}
% ------------------------------------ %
exploiting that the dynamics introduced by a control law $\kappa(\alpha)$ are described by 
% the probability transition matrix 
$\tens{P}_{\kappa}$.
This is a critical enabler for the control design 
as it allows the prediction of the invariant probability distribution $\vec{p}_{\kappa}^{\infty}$
by the eigenvector $\vec{p}_{\kappa}^{*1}$ associated 
with the dominant eigenvalue $\lambda^1_{\kappa}$ of $\tens{P}_{\kappa}$
(see appendix \ref{App:Sec:coCROM}).
Having determined $J_{\kappa}$ for all $\kappa$, the optimal control law is then given by 
% ------------ Equation -------------- %
\begin{equation}
 \kappa^{opt}(\alpha) = \mathrm{arg}\,\min_{b=\kappa(\alpha)} J_{\kappa}\quad\text{with}\quad J^{opt} = \min_{b=\kappa(\alpha)}  J_{\kappa}\, .
 \label{Eqn:OptimalControlLaw}
\end{equation}
% ------------------------------------ %
% that minimizes the objective function
% ------------ Equation -------------- %
% \begin{equation}
%  J^{opt} = \min_{b=\kappa(\alpha)}  J_{\kappa}\, . 
% \end{equation}
% ------------------------------------ %
% Finding the optimal control law from the set of all possible control laws in this discretized setting
% is a combinatorial optimization problem. 
% For a low number of possible control laws, an exhaustive search may be feasable.
% \textcolor{red}{TODO:Alternatives}

% Permutation is an ordered combination.
% Here combination with repetition

%% file: S4.tex
% ============================================================================================================== %
\subsection{Flow configuration and numerical simulation}
\label{Sec:FlowConfiguration}
\begin{figure}
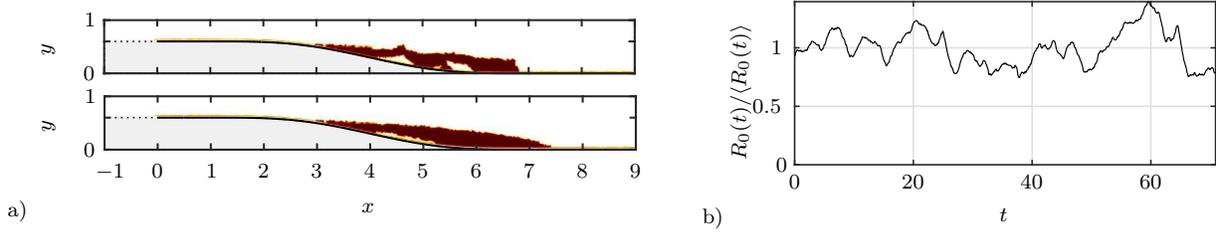

\centering
\begin{minipage}[c]{0.5\textwidth}
 \centering
 \psfrag{x}{$x$}
 \psfrag{y}{$y$}
 \psfrag{0}{$0$}
 \psfrag{0.6}{}
 \psfrag{1}{$1$}
 \centerline{\hspace{0.5cm}\includegraphics[width=0.98\textwidth,trim=0.4cm 1cm 0 0cm, clip=true]{Fig1-1}}
 \psfrag{-1}{$-1$}
 \psfrag{2}{$2$}
 \psfrag{3}{$3$}
 \psfrag{4}{$4$}
 \psfrag{5}{$5$}
 \psfrag{6}{$6$}
 \psfrag{7}{$7$}
 \psfrag{8}{$8$}
 \psfrag{9}{$9$}
 \centerline{a) \includegraphics[width=\textwidth,trim=0.2cm 0 0 0cm, clip=true]{Fig1-2}}
\end{minipage}
\hspace{1cm}
\begin{minipage}[c]{0.4\textwidth}
 \centering
 \psfrag{t}{$t$}
 \psfrag{R(t)}{\hspace{-0.7cm}$R_0(t)/\langle R_0(t)\rangle$}
 \psfrag{0}{$0$}
 \psfrag{0.5}{\hspace{-0.08cm}$0.5$}
 \psfrag{1}{\hspace{-0.08cm}$1$}
 \psfrag{10}{$10$}
 \psfrag{20}{$20$}
 \psfrag{30}{$30$}
 \psfrag{40}{$40$}
 \psfrag{50}{$50$}
 \psfrag{60}{$60$}
 \psfrag{70}{$70$}
 \centerline{b) \includegraphics[width=\textwidth,trim=0.25cm 0 0 0cm, clip=true]{Fig1-3}}
\end{minipage}
\caption{Recirculation area of (a, top) an instantaneous velocity snapshot and (a, bottom) the mean flow,
and (c) time series of the normalized recirculation area for the uncontrolled flow.}
\label{Fig:RecirculationArea}
\end{figure}

% In this section, the Navier-Stokes solver for the separating flow is outlined. 
The two-dimensional flow is described by a Cartesian coordinate system in which the 
location vector is denoted by $\vec{x} = (x, y)^T$ where $x$ 
is in flow direction and  $y$ is the direction perpendicular to $x$. 
The two-dimensional velocity vector is denoted by $\vec{u}(\vec{x}, t) := (u, v)^T$
where $u$ and $v$ are the velocities in $x$- and $y$-direction, respectively, 
and $t$ denotes the time. 
The pressure is represented by $P$. 
The non-dimensionalized Navier-Stokes and continuity equations are
% ------------ Equation -------------- %
\begin{eqnarray}
    \partial_t \vec{u} + \nabla\cdot (\vec{u}\,\vec{u}) &=& -\nabla P + \frac{1}{Re} \Delta \vec{u} + \vec{G}\,b,\notag\\
    \nabla\cdot \vec{u} &=& 0
\end{eqnarray}
% ------------------------------------ %
where $Re = U_{\infty} L/\nu$ is the Reynolds number and 
$\vec{G}$ is a steady local force field in $y$-direction. 
The function $b$ is the time-dependent control input amplitude
and has compact support in a circular region.
It is centered at $x = 1$ and the $y$-position is chosen such that the circular region is mostly inside the boundary layer.  
The computational domain $\Omega$ for the flow comprises
% ------------ Equation -------------- %
\begin{equation}
\Omega:= \left\{ (x,y):\, -1\leq x \leq 10,\,f(x)\leq y \leq 2.6  \right\}.
\end{equation}
% ------------------------------------ %
The domain is discretized as mixed Taylor-Hood elements \cite{Hood1974book} on an unstructured triangular mesh
comprising $8567$ nodes
% The grid consists of $8567$ nodes generating $16672$ triangles and has an 
with increased resolution around the leading edge, 
in the boundary layer and in the shear layer region. 
% The sponge region ranges from $x = 9$ to $x = 10$ in the streamwise direction and from $y = 0$ to $y = 2.6$ in the transverse direction. 
A quadratic finite-element method formulation is used to discretize the evolution equations
with no-slip boundary on the ramp and stress-free outflow. 
A detailed description of the solver can be found in \cite{Morzynski1987proc,Afanasiev2003phd}. 
% -------------- Figure -------------- %
% \begin{figure}
% \centering
% \psfrag{x}{$x$}
% \psfrag{y}{$y$}
% \psfrag{0}{\hspace{-0.05cm}$0$}
% \psfrag{0.6}{\hspace{-0.15cm}$0.6$}
% \psfrag{2.6}{\hspace{-0.15cm}$2.6$}
% \psfrag{-1}{\hspace{-0.2cm}$-1$}
% \psfrag{1}{\hspace{-0.05cm}$1$}
% \psfrag{2}{\hspace{-0.05cm}$2$}
% \psfrag{3}{\hspace{-0.05cm}$3$}
% \psfrag{4}{\hspace{-0.05cm}$4$}
% \psfrag{5}{\hspace{-0.05cm}$5$}
% \psfrag{6}{\hspace{-0.05cm}$6$}
% \psfrag{7}{\hspace{-0.05cm}$7$}
% \psfrag{8}{\hspace{-0.05cm}$8$}
% \psfrag{9}{\hspace{-0.05cm}$9$}
% \psfrag{10}{\hspace{-0.05cm}$10$}
% \includegraphics[width=0.65\textwidth, clip=true]{Figures/Grid_sponge}
% \caption{Computational domain of the numerical simulation. The smoothly contoured ramp starts at $x=0$. The sponge region starts at $x=9$.}
% \end{figure}
% ------------------------------------ %
A rectangular velocity profile $\vec{U}{\infty} := \vec{u}(x = −1, y) = (1, 0)^T$ is used as inflow. 
The numerical time step is $0.005$ and the sampling period of the snapshots is $20$, i.e. $\Delta t = 0.1$.
The topography of the smooth ramp is described by a polynomial shape of order $7$ \cite{Sommer1992vdi,Bao2004ast}.
Due to an adverse pressure gradient induced by the curvature of the ramp, 
the flow separates from the wall leading to a large recirculation area %directly behind the ramp
and the development of a convectively unstable free shear layer. 
It gives rise to the Kelvin-Helmholtz instability 
by which two-dimensional perturbations are spatially amplified 
eventually roll up into vortices \cite{Ho1984arfm}. 
% These shed with a certain frequency, are convected downstream, and eventually develop three-dimensional instabilities.
% Behind the separation point an open laminar separation bubble emerges.
% which
% downstream area is characterized by the unsteady vortex shedding originating from the Kelvin-Helmholtz instability. 
This recirculation area is characterized by fluid moving in the opposite direction of the flow. 
High pressure drag and low lift forces are associated with large recirculation areas on airfoils. 
% Hence, the goal of controlling laminar separation bubbles is generally the drag reduction and lift enhancement by, e.g.,
% mitigating the fluctuation energy or by decreasing the size of the recirculation area. 
In this study, the objective is to reduce the recirculation area in order to attenuate the pressure drag. 
The mean recirculation area $\langle R(t) \rangle$ is defined by
% as the temporal mean of the instantaneous recirculation area $R$
% ------------ Equation -------------- %
\begin{equation}
 \langle R(t) \rangle = \frac{1}{T_2-T_1}\int\limits_{T_1}^{T_2}\,\int\limits_{\Omega_R}\, H(- u(\vec{x}))(t)\,\mathrm d\vec{x}\,\mathrm dt 
\label{Eqn:Ramp2u:RecirculationArea}
\end{equation}
where $H$ denotes the Heaviside function,
$\Omega_R$ is the chosen region for evaluation,
and the limits for the temporal integration are chosen such that the transient is excluded,
i.e. $T_1 = 25 \approx 4.4\, T_{sh}$ and $T_2 - T_1 = 70 \approx 7.92\, T_{sh}$
with the shedding period $T_{sh} = 1/f_{sh}$ of the uncontrolled flow. 
% period of oscillation of the Kelvin-Helmholtz vortices for the uncontrolled flow. 
The estimation of the recirculation area is an approximation 
assuming that the recirculation area corresponds to those regions where the streamwise velocity
component is negative. 
The average cost function to assess the performance of the control is defined by
% ------------ Equation -------------- %
\begin{equation}
 J = \frac{\langle R(t)\rangle}{\langle R_0(t)\rangle} 
\label{Eqn:Ramp2u:ControlObjective}
\end{equation}
% ------------------------------------ %
normalized by the mean recirculation area $\langle R_0(t)\rangle$ of the uncontrolled flow. % denoted by the subscript '$0$'.
An instantaneous and mean plot of the recirculation area of the uncontrolled flow are shown in figure~\ref{Fig:RecirculationArea}.
% -------------- Figure -------------- %
% ------------------------------------ %

% For later reference, we present briefly the proper orthogonal decomposition (POD) \cite{Holmes2012book}.
% POD rests on a time or ensemble average denoted by $\langle \bullet \rangle$. 
% For a considered snapshot ensemble $\{ \vec{u}^m \}_{m=1}^M$, where $\vec{u}^m:=\vec{u}(t_m,\vec{x})$ at discrete times $t_m$, $m=1,\ldots,M$, 
% the conventional average is employed. 
% POD generalizes the Reynolds decomposition of a flow into 
% a mean flow $\vec{u}_0(\vec{x}):= \langle\vec{u}(\vec{x}, t)\rangle$ 
% and a fluctuation $\vec{u}'(\vec{x}, t)$.
% The POD expansion %\citep{Holmes2012book} 
% reads
% % ------------- Equation ------------- %
% \begin{equation}
%  \vec{u}'(\vec{x}, t) := \vec{u}(\vec{x}, t)-\vec{u}_0(\vec{x})
%  \approx \sum\limits_{i=1}^N\, a_i(t)\, \vec{u}_i(\vec{x})
% \end{equation}
% % ------------------------------------ %
% where $\vec{u}_i$ are the spatial POD modes 
% and $a_i:=(\vec{u}',\vec{u}_i)_{\Omega}$, $i=1,\ldots,N$,
% are the temporal mode coefficients.

% ============================================================================================================== %
% \subsection{Cluster and model identification}
\subsection{Control results}
% In this section, the cluster and model identification process is briefly summarized and the main control results are presented.
We consider feedback control laws $\kappa$ of the form presented in \eqref{Eqn:ControlLaw}
where the periodic excitation is turned on ($\chi_{\alpha}(\vec{a})=1$) or off ($\chi_{\alpha}(\vec{a})=0$)
depending on the prevailing cluster $\alpha$.
A suitable frequency for $\omega_p$ can be easily determined using open-loop periodic forcing and 
selecting the frequency for which the recirculation area has decreased most. 
For the considered flow simulation, 
this frequency has been determined as $f_{p}=0.45$.
Thus, the control law is based on the best periodic excitation 
exploiting that this frequency is known to be effective.
The system state $\vec{a} := [a_1,\ldots,a_N]^T$ is %therefore 
given by 
% the POD coefficients obtained from projecting the acquired snapshot onto the first $N_{pod} = 10$ modes
% corresponding to the best periodic forcing,
% ------------ Equation -------------- %
\begin{equation}
%  \hat{\vec{a}}(t_m) = 
 \vec{a}(t_m) = \int\limits_{\Omega} \vec{\Phi}_N^T\,\vec{u}^m(\vec{x})\,\mathrm d\vec{x}
 \label{Eqn:FeatureVector}
\end{equation}
% ------------------------------------ %
projecting the instantaneous velocity snapshot $\vec{u}^m$
onto the first $N_{pod}=10$ proper orthogonal decomposition \cite{Holmes2012book} (POD) modes $\vec{u}_i^{p}(\vec{x})$
constituting the columns of $\vec{\Phi}_N:= [\vec{u}_1^{p}, \ldots, \vec{u}_N^{p}]$.
These POD modes are computed from a snapshot ensemble sampled of a flow under periodic excitation with $f_p=0.45$.

\begin{figure}
\centering
\psfrag{B(t)}{$b(t)$}
\psfrag{t}{$t$}
\psfrag{-2}{\hspace{-0.1cm}$-2$}
\psfrag{-1}{$-1$}
\psfrag{0}{$0$}
\psfrag{1}{$1$}
\psfrag{2}{$2$}
\psfrag{50}{$50$}
\psfrag{100}{$100$}
\psfrag{150}{$150$}
% \psfrag{tau}{$\tau_i$}
\psfrag{tau}{\hspace{-0.5cm}$\tilde{\kappa}\sin(\omega_p\,t)$}
\includegraphics[width = 0.5\textwidth]{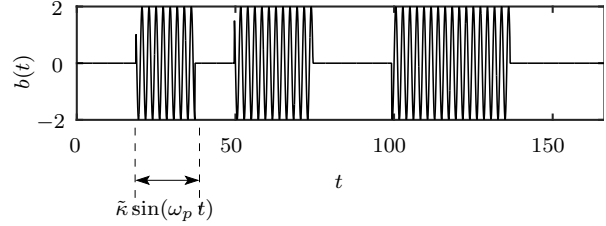}
\caption{Actuation signal for model identification switching between phases where $b=\tilde{\kappa}\,\sin(\omega_p\,t)$ and $b=0$.}
\label{Fig:ActuationSignal}
\end{figure}

\begin{figure}
\centering
$\vec{c}_1$\,\includegraphics[width=0.4\textwidth]{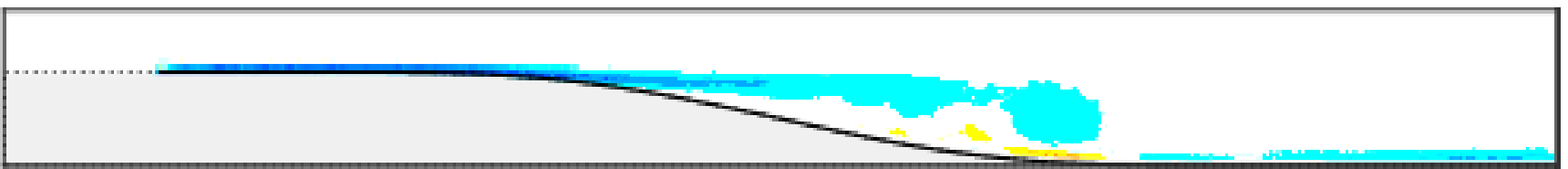}
\hspace{1cm}
$\vec{c}_6$\,\includegraphics[width=0.4\textwidth]{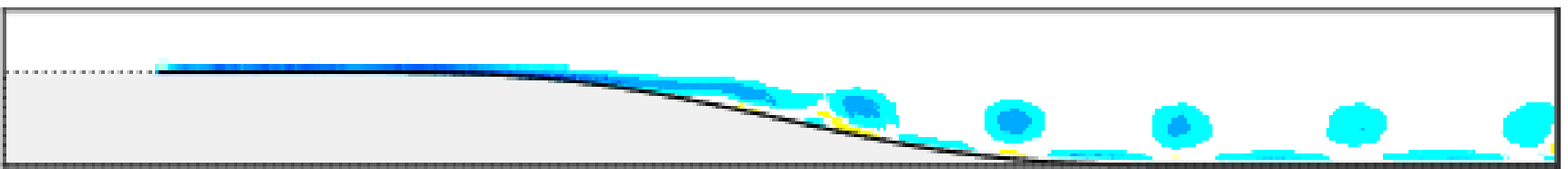}\\
$\vec{c}_2$\,\includegraphics[width=0.4\textwidth]{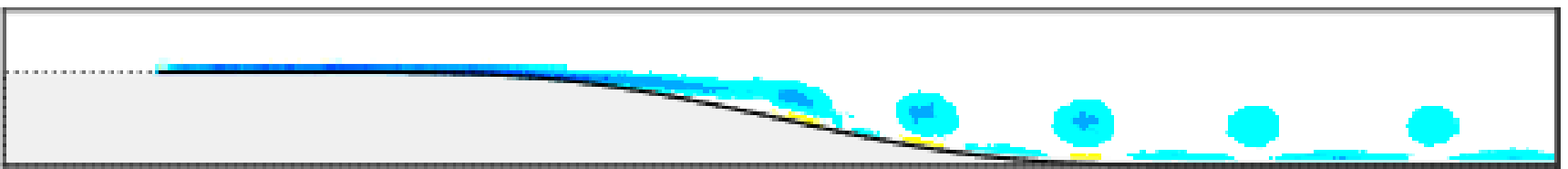}
\hspace{1cm}
$\vec{c}_7$\,\includegraphics[width=0.4\textwidth]{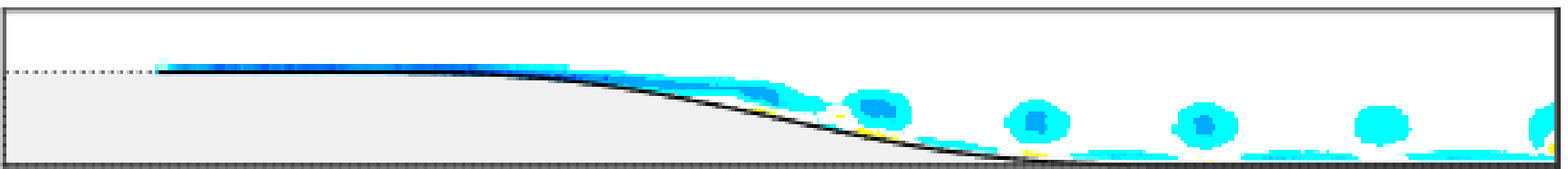}\\
$\vec{c}_3$\,\includegraphics[width=0.4\textwidth]{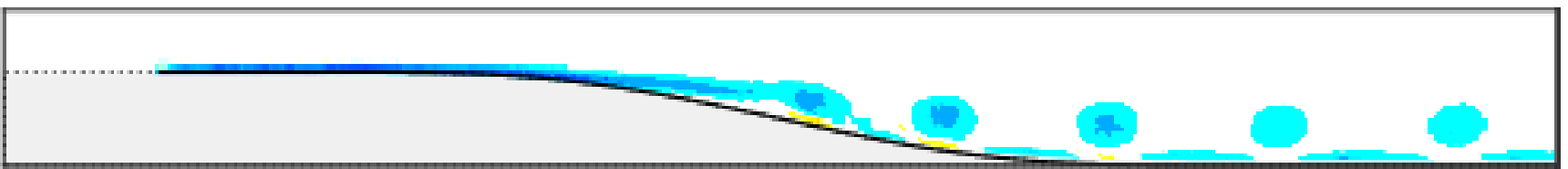}
\hspace{1cm}
$\vec{c}_8$\,\includegraphics[width=0.4\textwidth]{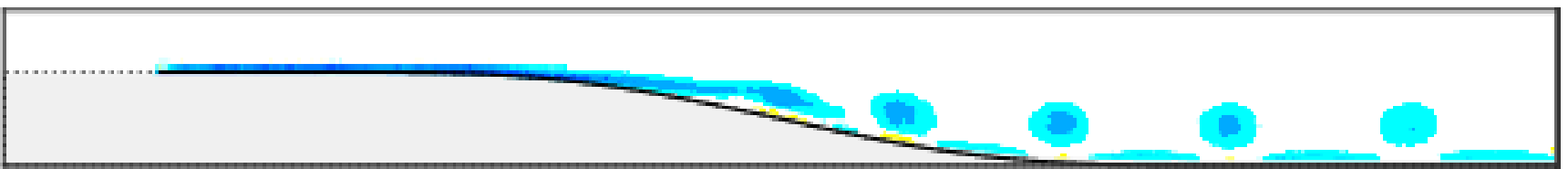}\\
$\vec{c}_4$\,\includegraphics[width=0.4\textwidth]{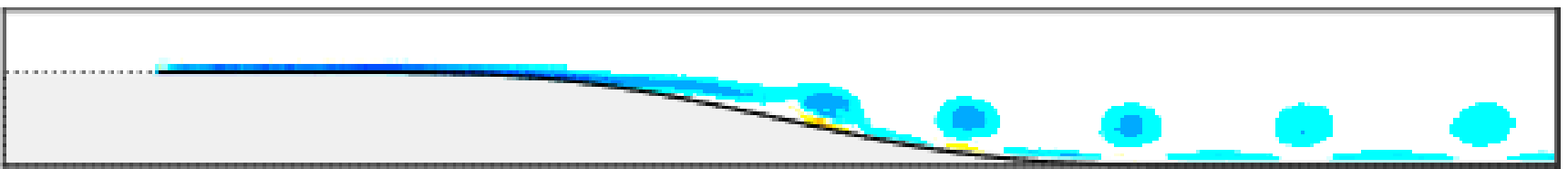}
\hspace{1cm}
$\vec{c}_9$\,\includegraphics[width=0.4\textwidth]{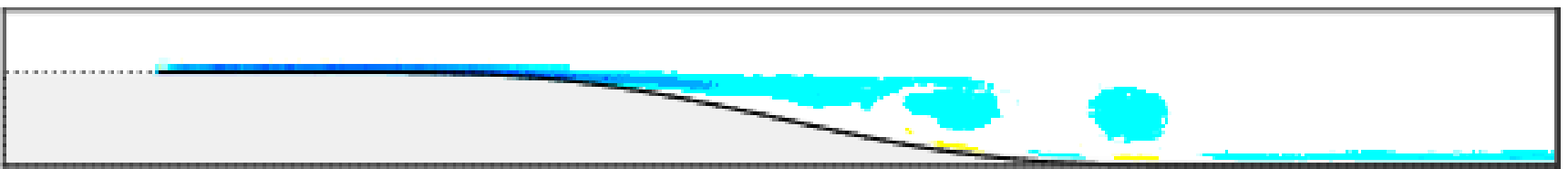}\\
$\vec{c}_5$\,\includegraphics[width=0.4\textwidth]{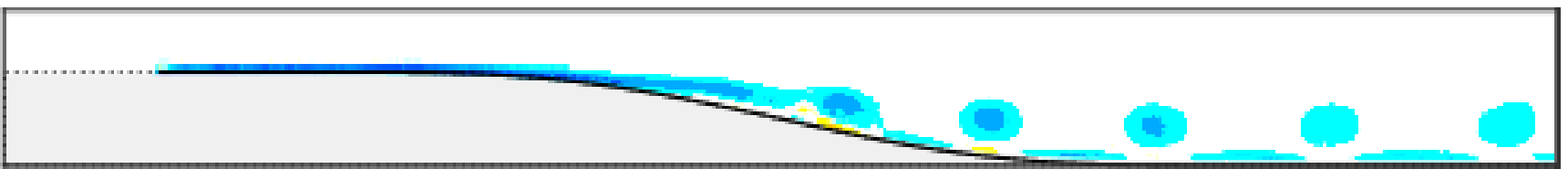}
\hspace{1cm}
$\vec{c}_{10}$\,\includegraphics[width=0.4\textwidth]{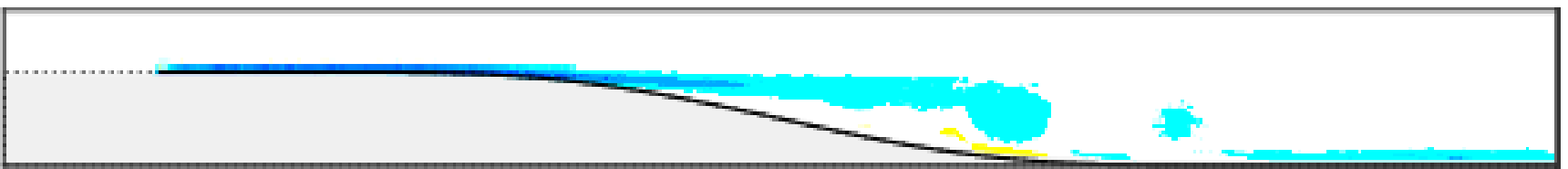}\\
\caption{Cluster vorticity centroids $\vec{c}_i(\vec{x}) = \nabla\times (1/n_i\,\sum_{\vec{a}^m\in\mathcal{A}_i}\,\vec{u}^m(\vec{x}))$ for each cluster $\mathcal{A}_i$.}
\label{Fig:Centroids}
\end{figure}

The data for the cluster and model identification is collected from applying the actuation signal 
(see figure~\ref{Fig:ActuationSignal}), 
which comprises time spans where the control is either turned on %, $\tau_i^{on}$, $i=2,4,6$, 
or off. %, $\tau_i^{off}$, $i=1,3,5,7$, 
% corresponding to the states $\chi_{\alpha}(\vec{a})$ can assume.
% -------------- Figure -------------- %
% ------------------------------------ %
The temporal signal of the POD coefficient vector $\vec{a}$ is computed from the acquired snapshot set 
according to \eqref{Eqn:FeatureVector}.
The state space is discretized by applying an unsupervised clustering algorithm (see App.~\ref{App:Sec:ClusterAnalysis}) 
to   the data ensemble $\{\vec{a}^m\}_{m=1}^M$ with the number of clusters $N_a=10$.
% The number of clusters is chosen to be $N_a=10$ to allow the evaluation of all possible control laws. 
The cluster centroids based on the vorticity of the snapshots belonging to each cluster
are displayed in figure~\ref{Fig:Centroids}. 
% -------------- Figure -------------- %
% ------------------------------------ %
Most centroids, $i=2,3,4,5,6,7,8$, represent the lock-in state when periodically exciting the flow.
The remaining centroids, $i=1,9,10$, are associated with the uncontrolled flow.
Transient states are not resolved. 

A local cost function $j(i)$ is associated with each cluster $\mathcal{A}_i$
% ------------ Equation -------------- %
\begin{equation}
 j(i) = \frac{1}{n_i}\sum\limits_{\vec{a}^m\in\mathcal{A}_i} R(t_m).
 \label{Eqn:ClusterCostArea}
\end{equation}
% ------------------------------------ %
in terms of the recirculation area averaged over the snapshots belonging to cluster $\mathcal{A}_i$.
The desirability of a particular cluster is thus represented 
by this cost taking into account the control objective (compare Sec.~\ref{Sec:FlowConfiguration}),
while the control input is not penalized here.
The control-dependent transition probabilities are computed according to \eqref{Eqn:Ulam_CTM} in App.~\ref{App:Sec:coCROM} 
and are used to construct the cluster transition matrices $\tens{P}_{\kappa}$
prescribing the dynamics under control law $\kappa$.
% Note that these matrices also include transients.

\begin{figure}
\centering
\psfrag{Noise}{Noise}
\psfrag{w}{$w$}
\psfrag{Control Input}{\hspace{-0.15cm}Control input}
\psfrag{b}{\hspace{-0.75cm}$b$}
\psfrag{b = pi(chi)}{$b = \kappa(\alpha)$}
\psfrag{Performance sensor}{Performance sensor}
\psfrag{z}{$J$}
\psfrag{Sensor Output}{Sensor output}
\psfrag{s}{\hspace{0.75cm}$\vec{s}$}
\psfrag{chi = f(s)}{$\alpha = \chi(\vec{s})$}
\psfrag{System}{System}
\psfrag{Controller}{Controller}
\includegraphics[width=0.65\textwidth]{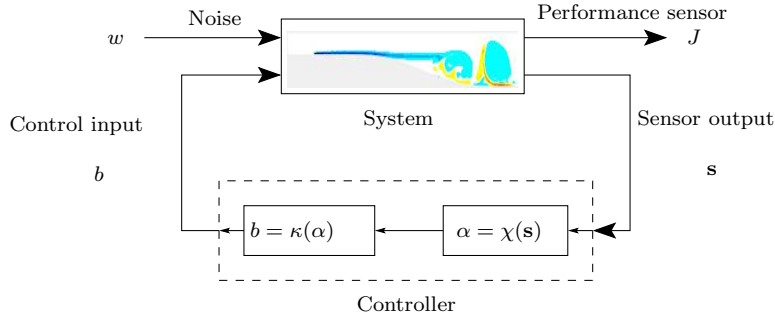}
\caption{Schematic of feedback control loop.}
\label{Fig:FeedbackLoop}
\end{figure}

\begin{figure}
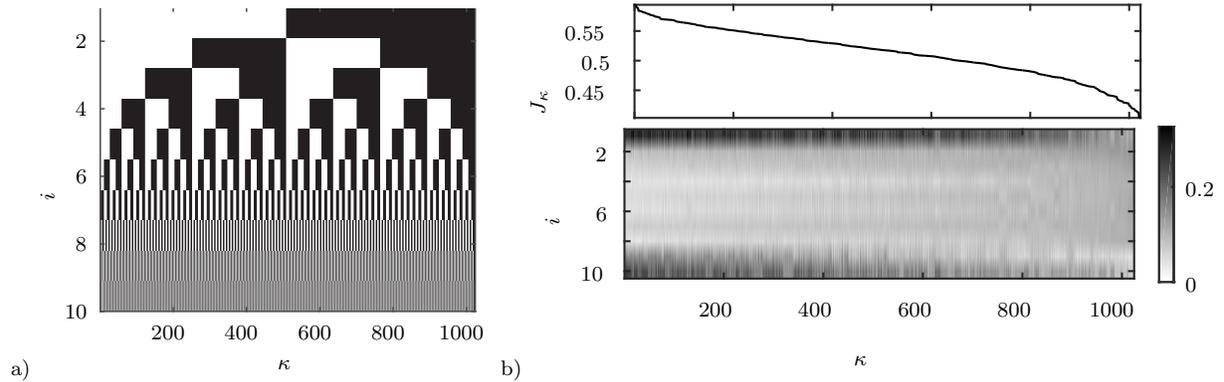

\centering
\begin{minipage}[b]{0.4\textwidth}
 \centering
 \psfrag{cluster}{$i$}
 \psfrag{control law}{\hspace{0.75cm}$\kappa$}
\psfrag{2}{\hspace{-0.05cm}$2$}
\psfrag{4}{\hspace{-0.05cm}$4$}
\psfrag{6}{\hspace{-0.05cm}$6$}
\psfrag{8}{\hspace{-0.05cm}$8$}
\psfrag{10}{\hspace{-0.05cm}$10$}
\psfrag{200}{\hspace{-0.05cm}$200$}
\psfrag{400}{\hspace{-0.05cm}$400$}
\psfrag{600}{\hspace{-0.05cm}$600$}
\psfrag{800}{\hspace{-0.05cm}$800$}
\psfrag{1000}{\hspace{-0.05cm}$1000$}
\centerline{a)\includegraphics[width=0.95\textwidth,clip=true]{Fig5-1}}
\end{minipage}
\hspace{0.05cm}
\begin{minipage}[b]{0.58\textwidth}
 \centering
\psfrag{0.45}{$0.45$}
\psfrag{0.5}{$0.5$}
\psfrag{0.55}{$0.55$}
\psfrag{500}{$500$}
\psfrag{expected costs}{\hspace{1.25cm}$J_{\kappa}$}
\hspace{-0.75cm}\includegraphics[width=0.93\textwidth,trim=0 1.15cm 0 1cm, clip=true]{Fig5-2}
\psfrag{0}{$0$}
\psfrag{0.2}{$0.2$}
\psfrag{2}{$2$}
\psfrag{6}{$6$}
\psfrag{10}{$10$}
\psfrag{200}{$200$}
\psfrag{400}{$400$}
\psfrag{600}{$600$}
\psfrag{800}{$800$}
\psfrag{1000}{$1000$}
\psfrag{cluster}{\hspace{0.5cm} $i$}
\psfrag{control law}{\hspace{1cm} $\kappa$}
\centerline{b)\includegraphics[width=\textwidth,clip=true]{Fig5-3}}
\end{minipage}
\caption{
Optimal control law:
(a) set of control laws $\kappa^{l}$, $l\in\{1,\ldots,1024\}$, 
(b, top) their average cost $J_{\kappa}$ (see \eqref{Eqn:AverageCostEigenvector}), and 
(b, bottom) the eigenvector $\vec{p}^{*1}$ % = [p_1^{*1},\ldots,p_i^{*1},\ldots,p_{10}^{*1}]$ 
corresponding to the dominant eigenvalue $\lambda_1$ 
of the associated transition matrix $\tens{P}_{\kappa}$.
The symbols $\square$ and $\blacksquare$ depict the control state 'off', i.e.\ $\chi_{\alpha}(\vec{a})= 0$, 
or 'on', i.e.\ $\chi_{\alpha}(\vec{a})= 1$, respectively.
Note that the control laws in both figures in (b) are sorted according to descending costs $J_{\kappa}$.
Thus, the best control law corresponds to the right most control law.}
\label{Fig:ControlLawEvaluation}
\end{figure}

% ============================================================================================================== %
% \subsection{Control results}
% In this section, the  results for all combinations of control laws are summarized.
The feedback control loop is displayed in figure~\ref{Fig:FeedbackLoop}.
% -------------- Figure -------------- %
% ------------------------------------ %
A sensor reading $\vec{s}$ is fed back to the controller in which 
first the prevailing cluster $\alpha$ is computed and then 
the control input $b$ is determined based on the control law $\kappa(\alpha)$.
Here, full-state information is assumed, i.e.\ $\vec{s}=\vec{a}$.
A realistic system is generally affected by noise which is neglected in this study.
A sensor measures the performance $J$ of the control law with regard to the control objective.
The set of control laws to be evaluated is shown in figure~\ref{Fig:ControlLawEvaluation}(b).
% -------------- Figure -------------- %
% ------------------------------------ %
The abscissa depicts the control laws $\kappa^l(\alpha)$ and the ordinate corresponds to the cluster index $i$, 
which is selected by the prevailing cluster $\alpha$.
The asymptotic probability vectors $\vec{p}_{\kappa}^{\infty}\,\forall\kappa$ 
originating from the dynamics prescribed by $\tens{P}_{\kappa}$ 
can be predicted by the corresponding eigenvector $\vec{p}_{\kappa}^{*,1}$ (see figure~\ref{Fig:ControlLawEvaluation}(b, bottom)).
The associated average cost \eqref{Eqn:AverageCostEigenvector} is displayed above in figure~\ref{Fig:ControlLawEvaluation}(b, top). 
The optimal control law $\kappa^{opt} = B_{0010101111}$ 
as determined by \eqref{Eqn:OptimalControlLaw} is the right most.
The left most probability vectors are clearly in favor of cluster $1$, $9$ and $10$,  
and have comparably low probabilities in the remaining clusters. 
For these control laws, the flow remains mostly in the clusters corresponding to the uncontrolled flow (compare figure~\ref{Fig:Centroids}).
In contrast, for those probability vectors on the right-hand side, the condition is reversed:
the probabilities of clusters $1$, $9$ and $10$ are much lower and those of the remaining clusters have increased.
Thus, these control laws direct the flow to clusters associated with smaller recirculation areas. 

\begin{figure}
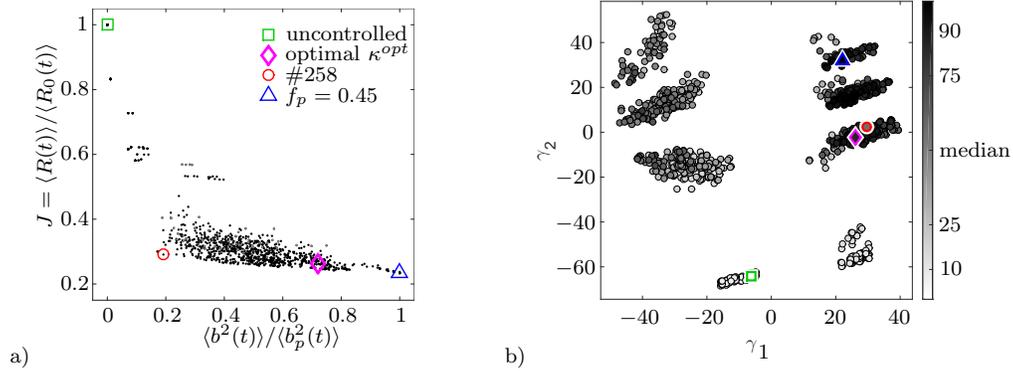

\centering
\psfrag{0}{$0$}
\psfrag{0.2}{$0.2$}
\psfrag{0.4}{$0.4$}
\psfrag{0.6}{$0.6$}
\psfrag{0.8}{$0.8$}
\psfrag{1}{$1$}
\psfrag{Recirculation Area}{$J = \langle R(t) \rangle / \langle R_0(t) \rangle$}
\psfrag{Input Energy}{$\langle b^2(t) \rangle / \langle b^2_p(t) \rangle$}
\psfrag{uncontrolled}{\hspace{-0.1cm}uncontrolled}
\psfrag{crom}{\hspace{-0.1cm}optimal $\kappa^{opt}$}
\psfrag{#258}{\hspace{-0.1cm}$\# 258$}
\psfrag{fp = 0.45}{\hspace{-0.1cm}$f_p=0.45$}
a)\;\raisebox{0.2cm}{\includegraphics[height=4.5cm,clip=true]{Fig6-1}}
\hspace{1cm}
\psfrag{0}{$0$}
\psfrag{20}{$20$}
\psfrag{40}{$40$}
\psfrag{-20}{\hspace{-0.1cm}$-20$}
\psfrag{-40}{\hspace{-0.1cm}$-40$}
\psfrag{-60}{\hspace{-0.1cm}$-60$}
\psfrag{10}{$10$}
\psfrag{25}{$25$}
\psfrag{median}{median}
\psfrag{75}{$75$}
\psfrag{90}{$90$}
b)\;\includegraphics[height=4.8cm,clip=true]{Fig6-2}
\caption{Control law evaluation: 
(a) Pareto diagram of control laws and 
(b) two-dimensional visualization of the control laws based on their similarity and 
colored by the percentile rank of their performance $J$.}
% their performance $J$ in terms of percentiles. }
\label{Fig:ParetoAnd2D}
\end{figure}

This analysis is based on the prediction of the models $\tens{P}_{\kappa}$. 
In addition, all control laws are evaluated in the numerical simulation.
The mean input energy is defined by
$\langle b^2(t) \rangle = \frac{1}{T_2-T_1}\int_{T_1}^{T_2}\,b^2(t)\,\mathrm dt$
% -------------- Equation ------------ %
% \begin{equation}
%  \langle b^2(t) \rangle = \frac{1}{T_2-T_1}\int\limits_{T_1}^{T_2}\,b^2(t)\,\mathrm dt.
% \end{equation}
% ------------------------------------ %
based on the applied actuation signal $b$ to assess the required control effort.
A Pareto diagram of the control results is shown in figure~\ref{Fig:ParetoAnd2D}(a).
% -------------- Figure -------------- %
% ------------------------------------ %
The axes of figure~\ref{Fig:ParetoAnd2D}(a) are normalized based on 
the mean recirculation area of the uncontrolled flow and the mean input energy of the open-loop periodic forcing.
The periodically forced flow has clearly the smallest recirculation area ($23\%$).
The optimal control law $\kappa^{opt}$ produces a slightly larger recirculation area ($26\%$).
Interestingly, the optimal control law yields a comparable $J$ while  
considerably decreasing the required input energy by $28\%$.
There is a trade-off: the recirculation area cannot be reduced without increasing the required input energy.
If these effects are weighted evenly, %balanced 
the best trade-off is achieved by control law $\# 258$ ($B_{0100000010}$) 
with a recirculation area corresponding to $29\%$ and
a reduction of the input energy by $81\%$.
Note that this control law can be attributed to the synchronization of the flow to recurring peaks 
in the actuation with frequency $f_p$.
Finally, the similarity of the control laws is analyzed.
In figure~\ref{Fig:ParetoAnd2D}(b), 
a two-dimensional plot is displayed where each circle corresponds to a particular control law $\kappa^l$.
The distance between these circles depicts their respective similarity as defined in \eqref{Eqn:SimilarityControlLaws} 
in App.~\ref{App:Sec:CLviz}.
The color of the circles depicts the percentile rank of $J$ associated with a particular control law, 
e.g.\ $90$ and higher correspond to the best $10\%$ of control laws.
The percentile rank is computed using the nearest rank method.
The control laws are arranged in several groups:
two lower bright groups corresponding to poorly performing control laws,
three large groups with mixed performance on the left-hand side,
and three groups of similar size on the right-hand side with majoritarily better performance.
Interestingly, $\kappa^{opt}$ and $\# 258$ belong to the same group 
while $\kappa^{opt}$ has a more similar performance with periodic forcing.
The grouping of the clusters is influenced by the state space discretization. % into clusters.
For example, the uncontrolled flow exhibits mainly clusters $1$, $9$, $10$ 
and with negligible probability clusters $2$ and $3$.
Any control law of the form $B_{000{\sf xxxxx}00}$, where ${\sf _x}$ can be any control value,
must be similar to $b(t)=0\,\forall t$ and perform like the uncontrolled flow,
and thus belong to the same group in figure~\ref{Fig:ParetoAnd2D}(b).
% If actuation is turned on solely in cluster $2$ or $3$, 
% the actuation might not be sufficient and 
% yields no significant improvement with respect to the uncontrolled flow.
% As a cons and are closely grouped to the uncontrolled flow.
Further analysis is required to clarify the exact origin of the grouping.

% with control law $\pi(\chi(t))$ that maps the cluster space $\hat{\vec{a}}(t)$ to a control command, 
% either $0$ ('off') or $1$ ('on').  with amplitude $\hat{b}$ and excitation frequency $f_p$.
% Sensor measurements are here the coefficients of a proper orthogonal decomposition (POD) 
% as obtained from projecting the instantenous velocity snapshots onto the first ten POD modes of the best open-loop actuation.

%% file: S5.tex
The present study proposes a cluster-based control strategy 
for the determination of optimal control laws for unsteady fluid flows.
%control of the ergodic measure on the attractor of unsteady fluid flows.
This framework builds upon a cluster-based reduced-order model (CROM) which 
translates high-dimensional, nonlinear dynamics into low-dimensional, probabilistic dynamics.
% CROM is closely aligned with closure schemes, 
% in which a stable fixed point represents the ergodic measure for the unsteady attractor in velocity space.
The control problem is formulated as a combinatorial optimization problem for the average cost.
% Finding the optimal control law from the set of all possible control laws in this discretized setting
% is a combinatorial optimization problem.
% For a low number of possible control laws, an exhaustive search may be feasable.
The ability to find the optimal control law in an unsupervised manner highlights 
the generic applicability of the framework to other dynamical systems. % without a priori knowledge of the system 

The approach is demonstrated for a separating flow over a smooth ramp with the aim to reduce the mean recirculation area.
An important observation %from evaluating the set of control laws 
is the trade-off 
between the recirculation area and the required control effort.
One cannot be decreased without increasing the other.
Intriguingly, while the number of clusters
is too low to resolve the transition process, 
considerably reductions in the input energy
can be achieved through the optimal control law %by applying the determined optimal control law
while yielding a similarly reduced recirculation area with respect to periodic excitation.
%performing similarly compared to the best periodic excitation. 
One particularly efficient control law is 
attributed to the synchronization of the flow to recurring peaks in the actuation. 
Similarly to sinusoidal forcing, the flow exhibits a lock-in state with the actuation frequency. 
As a consequence, the flow separates later with much smaller vortices shedding closely along the wall.

Discrete formulations like the proposed cluster-based control framework
%or similarly Markov decision processes (MDP) \cite{Howard1960book}
face the curse of dimensionality \cite{Bellman1961book}
where the high-dimensionality of the problem as a result of the discretization prevents an exhaustive search for the optimal solution.
%For MDPs, 
Model-free extensions like approximate dynamic programming exist 
% which learn the optimal solution 
% through iteratively updating the cost function or using function .
to approximate the optimal cost function, 
e.g. using function approximators \cite{Bertsekas2012book} or by successively improving the cost function through learning \cite{Sutton1998book}.
Alternatively, other optimization algorithms for the exploration of the solution space could be employed in order to circumvent this difficulty.
For example, genetic algorithm, an evolutionary optimization method classically used for parameter optimization \cite{Wahde2008book},
aims to find the optimal solution by
generating and evolving a set of candidate solutions based on the natural selection process. 
% A set of candidate control laws is generated and evolved using the natural selection process 
% converging eventually to a set which includes the most effective control laws.

The online-capability of control strategies is critical for their application in realistic configurations.
For a low number of sensors, e.g. of $\mathcal{O}(1)-\mathcal{O}(10^3)$,
the involved calculations when applying the control law are sufficiently fast.
If the flow state is based on velocity field measurements typically of $\mathcal{O}(10^5)-\mathcal{O}(10^7)$,
the clustering algorithm is applied in the POD space as similarly done in this study.
Then, the method will benefit from recent advances in compressed sensing \cite{Brunton2013} 
to find optimal sparse sensors in the high-dimensional velocity space  that determine the instantaneous cluster affiliation.

% An open problem generalization to continuous control inputs 
% reduce amount of data for identification of models

Flow control has a long tradition in scientific and engineering applications. 
We believe that recent advances in data science \cite{Brunton2015jcd,Brunton2015arx} and machine learning techniques \cite{Duriez2014aiaa,Gautier2015jfm}
for flow control will be transformative in the coming years. 
The cluster-based control framework, that is purely data-driven and determines optimal control laws in an unsupervised manner, 
contributes to this direction.
The proposed approach offers a promising new path for controlling the ergodic measure on the attractor 
% with built-in uncertainty and 
taking into account nonlinear actuation mechanisms.

% future control applications will benefit 
% 
% this future line ..
% examples
% 
% Paragraph 3: General comments on ensemble, liouville control, data-driven methods 
% 
% demonstrate method on
% 
% numerous applications / fields 
% 
% benefit

% \textcolor{red}{TODO:Alternatives}

%% file: appendix.tex
\section{Discrete domain decomposition using cluster analysis}
\label{App:Sec:ClusterAnalysis}
% Let $\mathcal{A}_i$, $i=1,\ldots,N_a$, represent a centroidal Voronoi tessellation of the state space such that 
% $\mathcal{A} = \cup_{i=1}^{N_a} \mathcal{A}_i$ with  $\mathcal{A}_i \cap \mathcal{A}_j = \emptyset$ for $i\neq j$,
% as obtained by, e.g., the clustering algorithm presented in App.~\ref{App:Sec:ClusterAnalysis}.
Cluster analysis is a part of machine learning and pattern recognition \cite{Bishop2007book}
which learns automatically from data.
The aim of cluster analysis is to find a hidden grouping among a given set of observations $\{\vec{a}^m\}_{m=1}^M$.
Here, k-means clustering \cite{Lloyd1982ieeetit} is employed
which  groups \textit{kinematically} similar flow states into a low number $N_a$ of clusters $\mathcal{A}_i$, $i=1,\ldots,N_a$, 
% Input is a sequence of observations $\{\vec{a}^m\}_{m=1}^M$ at discrete times $t_m$. 
% These observations are partitioned into $N_a$ clusters $\mathcal{A}_i$, $i=1,\ldots,N_a$
such that the similarity of observations in the same cluster is maximized 
while the similarity of observations belonging to different clusters shall be minimized.
Here, the {\it dissimilarity} between observations $\vec{a}^m$ and $\vec{a}^n$ is measured 
using the Euclidean distance
% ------------ Equation -------------- %
\begin{equation}
 D(\vec{a}^m,\vec{a}^n) := \vert\vert \vec{a}^m-\vec{a}^n\vert\vert_2.
 \label{Eqn:EuclideanDistance}
\end{equation}
% ------------------------------------ %
The cluster centroid $\vec{A}_i$ of $\mathcal{A}_i$ is defined as the average 
of observations belonging to the cluster 
$\vec{A}_i := \frac{1}{n_i}\,\sum_{\vec{a}^m\in\mathcal{A}_i}\,\vec{a}^m$ 
% ------------ Equation -------------- %
% \begin{equation}
%  \vec{A}_i := \frac{1}{n_i}\,\sum\limits_{\vec{a}^m\in\mathcal{A}_i}\,\vec{a}^m
%  \label{Eqn:Centroids}
% \end{equation}
% ------------------------------------ %
where $n_i$ is the total number of observations in cluster $\mathcal{A}_i$.
% The k-means algorithm partitions the data space $\mathcal{A}$ into $N_a$ centroidal Voronoi cells
% $\mathcal{A} = \cup_{i=1}^{N_a}\, \mathcal{A}_i$ 
% with  $\mathcal{A}_i \cap \mathcal{A}_j = \emptyset$ for $i\neq j$,
% which are defined as particular Voronoi cells for which the generating points of the
% Voronoi tessellation are equal to the mass centroids of the Voronoi regions \cite{Du1999siam}. 
The quality of the algorithm is monitored by the total cluster variance, 
$ J_{ca}\left( \vec{A}_1, \ldots, \vec{A}_{N_a} \right)
 = \sum_{i=1}^{N_a}\,\sum_{\vec{a}^m\in\mathcal{A}_i}\,
 \vert\vert \vec{A}_i-\vec{a}^m\vert\vert_{\mathcal{A}}^2$.
% ------------ Equation -------------- %
% \begin{equation}
%  J\left( \vec{A}_1, \ldots, \vec{A}_{N_a} \right)
%  = \sum\limits_{i=1}^{N_a}\,\sum\limits_{\vec{a}^m\in\mathcal{A}_i}\,
%  \vert\vert \vec{A}_i-\vec{a}^m\vert\vert_{\Omega}^2.
% \end{equation}
% ------------------------------------ %
The algorithm starts with an initial set of centroids and 
then iteratively improves them by minimizing the total cluster variance.
The set of optimal centroids is thus the solution of the optimization problem
% ------------ Equation -------------- %
\begin{equation}
 \vec{A}_1^{opt}, \ldots, \vec{A}_{N_a}^{opt}
  = \mathrm{arg}\min\limits_{\vec{A}_1, \ldots, \vec{A}_{N_a}} \,
  J\left( \vec{A}_1, \ldots, \vec{A}_{N_a} \right).
\end{equation}
% ------------------------------------ %
% The final result of the algorithm  
% corresponds to a centroidal Voronoi tessellation of the state space.
The reader is referred to \cite{Kaiser2014jfm} for more details.

\section{Control-oriented cluster-based reduced-order model}
\label{App:Sec:coCROM}
% In Sec.~\ref{Sec:Methodology}, the cluster-based control strategy is outlined 
% based on a discrete-state, discrete-time Markov model.
The propagator of the Markov model \eqref{Eqn:DiscreteStateDiscreteTimeMarkovModel}
for the coarse-grained dynamicas shall be directly inferred from data. 
% Let $\mathcal{A}_i$, $i=1,\ldots,N_a$, represent a centroidal Voronoi tessellation of the state space such that 
% $\mathcal{A} = \cup_{i=1}^{N_a} \mathcal{A}_i$ with  $\mathcal{A}_i \cap \mathcal{A}_j = \emptyset$ for $i\neq j$,
% as obtained by, e.g., the clustering algorithm presented in App.~\ref{App:Sec:ClusterAnalysis}.
% We assume a state space discretization $\mathcal{A}_i$, $i=1,\ldots,N_a$, 
% as obtained by the clustering algorithm presented in App.~\ref{App:Sec:ClusterAnalysis}.
% With each cluster $\mathcal{A}_i$ a probability $q_i$ can be associated that 
% the trajectory resides in this cluster.
% Input is a sequence of observations $\{\vec{a}^m\}_{m=1}^M$ at discrete times $t_m$. 
% We assume a discretization of the state space $\mathcal{A} = \cup_{i=1}^{N_a} \mathcal{A}_i$ obtained, 
% e.g., using the clustering algorithm outlined in Ap..~\ref{App:Sec:ClusterAnalysis}.  
% The probability vector $\vec{q} = [q_1,\ldots,q_K]^T$ with elements
% % ------------ Equation -------------- %
% \begin{equation}
%  q_i := \frac{n_i}{M},
% \end{equation}
% % ------------------------------------ %
% where $q_i > 0\;\forall i$,
% gives the probabilities of the trajectory residing in a particular cluster $\mathcal{A}_i$. 
% These probabilities are inferred from the data based on relative frequencies of cluster visits, 
% i.e.\ based on the number of observations $n_i$ belonging to cluster $\mathcal{A}_i$.
A multidimensional array $\tens{Q}\in\mathbb{R}^{N_a\times N_a\times 2}$ of control-dependent 
transition probabilities is constructed with elements
% ------------ Equation -------------- %
\begin{equation}
 \tens{Q}_{ijb} := \frac{\mathrm{card}\{\vec{a}^m \vert \vec{a}^m \in\mathcal{A}_j\text{ and } 
 \vec{a}^{m+1} \in\mathcal{A}_i \text{ and } b^m\}}{\mathrm{card}\{ \vec{a}^m \in \mathcal{A}_j \}}
 = \mathrm{Prob}\left( \alpha^{t+1} \vert \alpha^t, b^t \right)
 \label{Eqn:Ulam_CTM}
\end{equation}
% ------------------------------------ %
where $\mathrm{card}$ denotes cardinality. 
The element $\tens{Q}_{ijb}$ constitutes the conditional probability 
that at time $t+1$ the trajectory is in cluster $\mathcal{A}_i$ under the condition 
that at the previous time step $t$ the trajectory was in cluster $\mathcal{A}_j$ and control $b$ was applied.
Array $\tens{Q}$ is directly inferred from data based on the relative frequencies of cluster transitions.
The control-oriented cluster transition matrix (CTM) $\tens{P}_{\kappa}$ for a particular control law $\kappa$ 
is constructed from the data array $\tens{Q}_{ijb}$ as
% ------------ Equation -------------- %
\begin{equation}
 \tens{P}_{\kappa} := \left[ \vec{q}_{1\,\kappa(1)}\;\cdots\; \vec{q}_{j\,\kappa(j)} \;\cdots\;\vec{q}_{N_a\,\kappa(N_a)} \right].
\end{equation}
% ------------------------------------ %
with $\vec{q}_{j\kappa(j)} = [\tens{Q}_{1j\kappa(j)},\ldots,\tens{Q}_{N_aj\kappa(j)} ]^T$
by selecting the columns $\vec{q}_{j\kappa(j)}$ specified by the control law $\kappa$. 
% For instance, the transition matrix prescribing the dynamics following control law $\kappa^{386}=B_{0110000010}$ \eqref{Eqn:ControlLaw386}
% is 
% % ------------ Equation -------------- %
% \begin{equation}
%  \tens{P}_{\kappa^{386}} := \left[ \vec{q}_{1\,0}\;\vec{q}_{2\,1}\;
% 				   \vec{q}_{3\,1}\;\vec{q}_{4\,0}\;
% 				   \vec{q}_{5\,0}\;\vec{q}_{6\,0}\;
% 				   \vec{q}_{7\,0}\;\vec{q}_{8\,0}\;
% 				   \vec{q}_{9\,1}\;\vec{q}_{10\,0} \right].
% \end{equation}
% % ------------------------------------ %
% All elements of a CTM $\tens{P}$ are non-negative, i.e.\ $\tens{P}_{ij}\geq 0\; \forall\; i,j$. 
% The elements of each column sum up to unity, i.e. $\sum_{i=1}^{N_a}\,\tens{P}_{ij} = 1\; \forall\; j$, 
% i.e.\ the total probability for all transitions under a particular control command is one. 
% % The CTM is identified as the one-step transition probability matrix.
% This property preserves the normalization condition of the probability vector.
% The cluster probability vector with probabilities $p_k^t$ to be in cluster $\mathcal{C}_k$ at time $t$ 
% is denoted by $\vec{p}^t = [p_1^t,\ldots,p_K^t]^T$ 
% which has non-negative probabilities, i.e. $p_k^t \geq 0$, and
% fulfils the normalization condition $\sum_{k=1}^K\, p_k^t = 1$ for each time-step $t$.
% The evolution of the cluster probability vector can be described as follows. 
% Let $\vec{p}_0$ be the initial probability distribution. 
In the following, the temporal evolution of a general cluster probability vector $\vec{p} = [p_1,\ldots,p_{N_a}]^T$ is pursued.
Having an initial probability distribution $\vec{p}^0$,
the cluster probability vector at time $t$ is compactly given by %$\vec{p}^{t} = \tens{P}_{\kappa}^t\,\vec{p}^{0}$ 
% ------------ Equation -------------- %
\begin{equation}
 \vec{p}^{t} = \tens{P}_{\kappa}^t\,\vec{p}^{0},
 \label{Eqn:DTMC_from_p0}
\end{equation}
% ------------------------------------ %
where the dynamics are prescribed by $\tens{P}_{\kappa}$ following a particular control law $\kappa$.
The cluster probability vector has non-negative probabilities, i.e. $p_i^t \geq 0$, and
fulfils the normalization condition $\sum_{i=1}^{N_a}\, p_i^t = 1$ for each timestep $t$.
The long-term behaviour can be studied by powers of the CTM as defined in \eqref{Eqn:DTMC_from_p0}.
The asymptotic probability distribution is obtained by
% ------------ Equation -------------- %
\begin{equation}
 \vec{p}^{\infty} := \lim\limits_{t\rightarrow\infty} \tens{P}_{\kappa}^{t}\,\vec{p}^{0}.
\end{equation}
% which describes the long-term behavior of the system.
If $\vec{p}^t$ converges to a unique, stationary probability vector, 
the system can said to be \textit{ergodic}, in the sense that it will be probabilistically reproducable:
regardless of the initial region of state space in which it is sampled, 
the ensemble mean will converge in the infinite-time limit to the time mean.
Each propagator $\tens{P}_{\kappa}$ defines a time-homogeneous Markov chain with well-known
properties \cite{Meyer2000book}:
(i) The propagator $\tens{P}_{\kappa}$ is a stochastic matrix with 
non-negative elements, i.e.\ $\tens{P}_{ij}\geq 0\; \forall\; i,j$.
The elements of each column sum up to unity, i.e. $\sum_{i=1}^{N_a}\,\tens{P}_{ij} = 1\; \forall\; j$.
These properties preserve the normalization condition of the probability vector.
(ii) The sequence of probability vectors $\vec{p}^t$, $t = 0, 1, 2, \ldots$, has no long-term memory.
The state at iteration $t+1$ only depends on the $t$th state and not on any previous iterations.
(iii) The absolute values of all eigenvalues of this matrix do not exceed unity. 
This excludes a diverging vector sequence.
(iv) It exists an eigenvalue $\lambda_1( \tens{P}_{\kappa} ) = 1$ with algebraic multiplicity $1$ and all other
eigenvalues satisfy $\vert \lambda_i ( \tens{P}_{\kappa} )\vert < 1$ for $i = 2, \ldots, N_a$. 
This is a consequence of the Perron-Frobenius theory for non-negative matrices \cite{Meyer2000book}. 
The eigenvector $\vec{p}^{*1}$
associated with the dominant eigenvalue $\lambda_1( \tens{P}_{\kappa} )$ fulfils 
the fixed-point equation $\tens{P}_{\kappa}\,\vec{p}^{*1} = \vec{p}^{*1}$.
Since $\vert\lambda_i( \tens{P}_{\kappa} )\vert < 1$ for $i=2,\ldots, N_a$, 
the vector $\vec{p}^{*1}$ is the only one that survives ininite iterations.
Mathematically, the stationary probability vector $\vec{p}^{\kappa}$
must be identical with the eigenvector $\vec{p}^{*1}$ associated with the
dominant eigenvalue $\lambda_1 = 1$, and thus is a fixed point to \eqref{Eqn:DTMC_from_p0} for any $t$ . 
If, however, $\vec{p}^{\infty}$ is oscillatory or non-stationary, 
the system will not be probabilistically reproducible, 
displaying a more complicated connection between the initial sampling region and its convergence properties. 

\section{Visualization of control laws}
\label{App:Sec:CLviz}
% For a better assessment of the available information obtained from evaluating the set of control laws, e.g. how much the control space is explored, 
% we seek to visualize the control laws in the sense that similar control laws shall lie closely to each other. %gather together.
% For a better assessment of the control laws and their respective performance, 
% we seek to visualize their similarity. 
% For this purpose, a distance matrix $\tens{D}$ with elements
For the purpose of visualizing the similarity of the control laws, a distance matrix $\tens{D}$
% % ------------ Equation -------------- %
\begin{equation}
 \tens{D}_{ij} = \sqrt{ \frac{1}{2}\,\sum\limits_{t=1}^{T}\, \left(b_i(\vec{s}_i(t)) - b_j(\vec{s}_i(t))\right)^2
	       + \frac{1}{2}\,\sum\limits_{t=1}^{T}\, \left(b_i(\vec{s}_j(t)) - b_j(\vec{s}_j(t))\right)^2}
	       \label{Eqn:SimilarityControlLaws}
\end{equation}
% % ------------------------------------ %
is defined.
The $b_i(\vec{s}_i(t)):= \kappa^i(\vec{s}_i(t))$ is the time series of the control input  
based on sensor readings $\vec{s}_i$ 
when applying control law $\kappa^i$. 
%which are obtained when applying this particular control law.
The time series $b_i(\vec{s}_j(t)):= \kappa^i(\vec{s}_j(t))$
is obtained from evaluating $\kappa^i$ using sensor readings $\vec{s}_j$
which are collected when $\kappa^j$ was applied. 
This permutation is incorporated
% The evaluation of  sampled while applying other control laws $\kappa_j$ is incorporated
to ensure the symmetry of $\tens{D}$. %and zero-valued diagonal entries
Note that $\vec{s} = \vec{a}$ in the case of full-state information.

A simple method that optimally preserves the control laws' pointwise distances
in a least-mean-square-error sense is {\it multidimensional scaling} (MDS) 
\cite{Mardia1979book,Cox2000book}.
For a given distance matrix according to a (possibly non-Euclidean) distance metric, 
MDS aims to find corresponding points in a low-dimensional subspace 
so that the distances between the points are preserved. 
In particular, a two-dimensional subspace denoted by $\gamma_1$ and $\gamma_2$ 
for visualization purposes is of interest.
% This is referred to as classical scaling. 
The solution can vary in terms of a translation, a rotation and reflections. 
In the case where the distance is measured via the Euclidean metric, 
this method coincides with the POD, 
and the mean is at the origin and the axes are the POD eigenvectors \cite{Cox2000book}.

%% file: 2016_tcfd_KAISER.bbl
\begin{thebibliography}{10}
\providecommand{\url}[1]{{#1}}
\providecommand{\urlprefix}{URL }
\expandafter\ifx\csname urlstyle\endcsname\relax
  \providecommand{\doi}[1]{DOI~\discretionary{}{}{}#1}\else
  \providecommand{\doi}{DOI~\discretionary{}{}{}\begingroup
  \urlstyle{rm}\Url}\fi

\bibitem{Afanasiev2003phd}
Afanasiev, K.: Stabilit\"atsanalyse, niedrigdimensionale modellierung und
  optimale kontrolle der kreiszylinderumstr\"omung (trans.: Stability analysis,
  low-dimensional modeling, and optimal control of the flow around a circular
  cylinder).
\newblock Ph.D. thesis, Fakult\"at Maschinenwesen, Technische Universit\"at
  Dresden (2003)

\bibitem{Bao2004ast}
Bao, F., Dallmann, U.C.: Some physical aspects of separation bubble on a
  rounded backward-facing step (physikalische ph\"nomene von abl\"seblasen an
  einer abgerundeten zur\"ckspringenden stufe).
\newblock Aerospace Science and Technology \textbf{8}, 83--91 (2004)

\bibitem{Bellman1961book}
Bellman, R.E.: Adaptive {C}ontrol {P}rocesses.
\newblock Princeton University Press, New York (1961)

\bibitem{Bertsekas2012book}
Bertsekas, D.P.: Dynamic Programming and Optimal Control, Vol. II, 4th edn.
\newblock Athena Scientific (2012)

\bibitem{Bishop2007book}
Bishop, C.M.: {P}attern {R}ecognition and {M}achine {L}earning.
\newblock Springer, New York (2007)

\bibitem{Bollt2013book}
Bollt, E.M., Santitissadeekorn, N.: {A}pplied and {C}omputational {M}easurable
  {D}ynamics.
\newblock SIAM (2013)

\bibitem{Brockett2012}
Brockett, R.: Notes on the control of the {L}iouville equation.
\newblock In: P.~Cannarsa, J.M. Coron (eds.) Control of {P}artial
  {D}ifferential {E}quations. Cetraro, Italy 2010. Springer-Verlag, Berlin
  Heidelberg (2012)

\bibitem{Brockett2003ieee}
Brockett, R.W.: Minimizing {A}ttention in a {M}otion {C}ontrol {C}ontext.
\newblock Proceedings of the 42nd IEEE Conference on Decision and Control
  \textbf{3349--3352} (2003).
\newblock Maui, Hawaii USA

\bibitem{Brockett2010proc}
Brockett, R.W.: On the control of a flock by a leader.
\newblock In: Proceedings of the Steklov Institute of Mathematics, vol. 268,
  pp. 49--57 (2010)

\bibitem{Brunton2013}
Brunton, B.W., Brunton, S.L., Proctor, J.L., Kutz, J.N.: Optimal sensor
  placement and enhanced sparsity for classification.
\newblock arXiv: 1310.2417  (2015)

\bibitem{Brunton2015amr}
Brunton, S.L., Noack, B.R.: Closed-loop turbulence control: {P}rogress and
  challenges.
\newblock Appl. Mech. Rev. \textbf{67}(5), 050,801:01--48 (2015)

\bibitem{Brunton2015arx}
Brunton, S.L., Proctor, J.L., Kutz, J.N.: Discovering governing equations from
  data: Sparse identification of nonlinear dynamical systems.
\newblock arXiv: 1509.03580  (2015)

\bibitem{Brunton2015jcd}
Brunton, S.L., Proctor, J.L., Tu, J.H., Kutz, J.N.: Compressive sampling and
  dynamic mode decomposition.
\newblock To appear in J.\ Comp.\ Dynamics  (2015)

\bibitem{Cox2000book}
Cox, T.F., Cox, M.A.A.: Multidimensional {S}caling, \emph{Monographs on
  Statistics and Applied Probability}, vol.~88, 2nd edn.
\newblock Chapman and Hall (2000)

\bibitem{Duriez2014aiaa}
Duriez, T., Parezanovic, V., Laurentie, J.C., Fourment, C., Delville, J.,
  Bonnet, J.P., Cordier, L., Noack, B.R., Segond, M., Abel, M.W., Gautier, N.,
  Aider, J.L., Raibaudo, C., Cuvier, C., Stanislas, M., Brunton, S.L.:
  Closed-loop control of experimental shear layers using machine learning
  (invited).
\newblock AIAA Paper  (2014).
\newblock 7th AIAA Flow Control Conference, Atlanta, Georgia

\bibitem{Froyland2001}
Froyland, G.: Extracting dynamical behavior via markov models.
\newblock In: A.I. Mees (ed.) Nonlinear Dynamics and Statistics, pp. 281--321.
  Birkh\"auser Boston (2001)

\bibitem{Gautier2015jfm}
Gautier, N., Aider, J.L., Duriez, T., Noack, B.R., Segond, M., Abel, M.W.:
  Closed-loop separation control using machine learning.
\newblock Journal of Fluid Mechanics \textbf{770}, 242--441 (2015)

\bibitem{Ho1984arfm}
Ho, C.M., Huerre, P.: Perturbed free shear layers.
\newblock Ann.\ Rev.\ Fluid Mech. \textbf{16}, 365--424 (1984)

\bibitem{Holmes2012book}
Holmes, P., Lumley, J.L., Berkooz, G., Rowley, C.W.: Turbulence, {C}oherent
  {S}tructures, {D}ynamical {S}ystems and {S}ymmetry, 2nd paperback edn.
\newblock Cambridge University Press, Cambridge (2012)

\bibitem{Hood1974book}
Hood, P., Taylor, C.: Finite Element Methods in Flow Problems, chap.
  Navier–Stokes equations using mixed interpolation, pp. 121--132.
\newblock University of Alabama in Huntsville Press (1974)

\bibitem{Hopf1952jrma}
Hopf, E.: Statistical hydromechanics and functional analysis.
\newblock J.\ Rat.\ Mech.\ Anal. \textbf{1}, 87--123 (1952)

\bibitem{Iverson1962book}
Iversion, K.E.: A {P}rogramming {L}anguage, 2nd edn.
\newblock John Wiley \& Sons Inc (1962)

\bibitem{Kaiser2014jfm}
Kaiser, E., Noack, B.R., Cordier, L., Spohn, A., Segond, M., Abel, M.,
  Daviller, G., \"Osth, J., Krajnovi\'c, S., Niven, R.K.: Cluster-based
  reduced-order modelling of a mixing layer.
\newblock J.\ Fluid Mech. \textbf{754}, 365--414 (2014)

\bibitem{Lasota1994book}
Lasota, A., Mackey, M.C.: {C}haos, {F}ractals, and {N}oise, 2nd edn.
\newblock Springer New York (1994)

\bibitem{Lloyd1982ieeetit}
Lloyd, S.: Least squares quantization in {PCM}.
\newblock IEEE Trans. Inform. Theory \textbf{28}, 129--137 (1956).
\newblock Originally as an unpublished Bell laboratories Technical Note (1957)

\bibitem{Lorenz1963jas}
Lorenz, E.N.: Deterministic nonperiodic flow.
\newblock J.\ Atm.\ Sci. \textbf{20}, 130--141 (1963)

\bibitem{Majumdar2014ijrr}
Majumdar, A., Vasudevan, R., Tobenkin, M.M., Tedrake, R.: Convex {O}ptimization
  of {N}onlinear {F}eedback {C}ontrollers via {O}ccupation {M}easures.
\newblock International Journal of Robotics Research \textbf{33}, 1209--1230
  (2014)

\bibitem{Mardia1979book}
Mardia, K.V., Kent, J.T., Bibby, J.M.: Multivariate {A}nalysis.
\newblock Academic Press (1979)

\bibitem{Meyer2000book}
Meyer, C.D.: Matrix {A}nalysis and {A}pplied {L}inear {A}lgebra.
\newblock Society for Industrial and Applied Mathematics (2000)

\bibitem{Morzynski1987proc}
Morzy\'nski, M.: Numerical solution of navier-stokes equations by the finite
  element method.
\newblock In: Proceedings of SYMKOM 87, Compressor and Turbine Stage Flow Path
  -- Theory and Experiment, pp. 119--128 (1987)

\bibitem{Munowitz1987jcp}
Munowitz, M., Pines, A., Mehring, M.: Multiple-quantum dynamics in {N}{M}{R}:
  {A} directed walk through {L}iouville space.
\newblock J.\ Chem.\ Phys. \textbf{86}, 3172--3182 (1987)

\bibitem{Noack2012jfm}
Noack, B.R., Niven, R.K.: Maximum-entropy closure for a {G}alerkin system of
  incompressible shear flow.
\newblock J.\ Fluid Mech. \textbf{700}, 187--213 (2012)

\bibitem{Sommer1992vdi}
Sommer, F.: Mehrfachl\"osungen bei laminaren str\"omungen mit druckinduzierter
  abl\"osung: eine kuspen-katastrophe (transl.: Multiple solutions of laminar
  flows with pressure induced separation: a cusp catastrophe).
\newblock Tech. Rep. 7:206, Fortschrittberichte VDI, VDI Verlag, D\"usseldorf
  (1992)

\bibitem{Sutton1998book}
Sutton, R.S., Barto, A.G.: Reinforcement {L}earning: {A}n {I}ntroduction.
\newblock MIT Press, Cambridge, MA (1998)

\bibitem{Ulam1964book}
Ulam, S.: Problems in {M}odern {M}athematics.
\newblock Interscience (1964)

\bibitem{Wahde2008book}
Wahde, M.: Biologically {I}nspired {O}ptimization {M}ethods: {A}n
  {I}ntroduction.
\newblock WIT Press (2008)

\end{thebibliography}
